\documentclass{aastex}

\begin{document}
 
\title{What are the Hot R Coronae Borealis Stars?}
 
\author{Orsola De Marco$^{1}$, 
Geoffrey C. Clayton$^{2}$, 
F. Herwig$^{3}$, 
D.L. Pollacco$^{4}$ 
J.S. Clark$^5$
\& David Kilkenny$^{6}$}
  
\altaffiltext {1}{Department of Astrophysics,
American Museum of Natural History,
New York, NY 10024-5192; Electronic mail: orsola@amnh.org}

\altaffiltext{2}{Department of Physics and Astronomy, Louisiana State University, 
Baton Rouge, LA 70803; 
Electronic mail: gclayton@fenway.phys.lsu.edu}

\altaffiltext{3}{University of Victoria, Box 3055, Victoria B.C., V8W 3P6, Canada;
Electronic mail: fherwig@uvastro.phys.uvic.ca} 

\altaffiltext{4}{Department of Pure and Applied Physics, Queen's University Belfast, 
University Road, Belfast BT7 1NN, UK; Electronic mail: D.Pollacco@Queens-Belfast.AC.UK}

\altaffiltext{5}{Department of Physics \& Astronomy, University College London,  
Gower Street, London WC1E 6BT, UK; Electronic mail: jsc@star.ucl.ac.uk             }

\altaffiltext{6}{South African Astronomical Observatory, P.O. Box 9, Observatory 7935,
South Africa; Electronic mail: dmk@da.saao.ac.za}

\begin{abstract}
We investigate the evolutionary status of four stars: V348 Sgr, DY Cen
and MV Sgr in the Galaxy and HV~2671 in the LMC. These stars have in
common random deep declines in visual brightness which are
characteristic for R Coronae Borealis (RCB) stars.  RCB stars are typically
cool,
hydrogen deficient supergiants.  The four stars studied in this paper
are hotter (T$_{\rm eff}$ = 15-20~kK) than the majority of RCB stars
(T$_{\rm eff}$ = 5000-7000~K). Although these are commonly grouped
together as the \emph{hot RCB stars} they do not necessarily share a
common evolutionary history.  We present new observational data and an
extensive collection of archival and previously-published data which is
reassessed to ensure internal consistency.  We find temporal variations
of various properties on different time scales which will eventually
help us to uncover the evolutionary history of these objects. DY Cen and
MV Sgr have typical RCB helium abundances which excludes any currently known
post-AGB evolutionary models. Moreover, their carbon and nitrogen abundances
present us with further problems for their interpretation. V348 Sgr and HV~2671 are in
general agreement with a born-again post-AGB evolution and their
abundances are similar to Wolf-Rayet central stars of PN. The three
Galactic stars in the sample have circumstellar nebulae which produce
forbidden line radiation (for HV~2671 we have no information). V348 Sgr
and DY Cen have low density, low expansion velocity nebulae (resolved
in the case of V348 Sgr), while MV Sgr has a higher density, higher
expansion velocity nebula. All three stars on the other hand have
split emission lines which indicate the presence of an equatorial bulge but
{\it not} of a Keplerian disk. In
addition, the historical light-curves for the three Galactic hot RCB
stars, show evidence for a significant fading in their maximum-light
brightnesses of $\sim$1~mag over the last 70~yr. From this we deduce
that their effective temperature increased by a few thousand degrees.
If V348 Sgr is a born-again star, as we presume, this means that 
the star is returning from the born-again
AGB phase to the central star of PN phase.  Spectroscopically, no dramatic change is
observed over the last 50 years for V348 Sgr and MV Sgr.  However,
there is some evidence that the winds of V348 Sgr and DY Cen have
increased in strength in the last decade.  HV~2671, located in the LMC
has not been analyzed in detail, but at 5-\AA\ resolution is almost
identical to V348 Sgr. Through the bolometric correction derived for
V348 Sgr and the known distance, we can estimate the absolute V
magnitude of HV~2671 ($M_V$ = --3.0~mag) and its bolometric luminosity
($\sim$6000~L$_\odot$).
\end{abstract}

\section{Introduction}
\label{sec:introduction}

The R Coronae Borealis (RCB) stars are hydrogen-deficient supergiants,
characterized by spectacular drops in visual brightness which occur at random
intervals. RCB stars are a very rare phenomenon with only
about 35 known in the Galaxy.
The atmospheres of these stars are composed mostly of helium with carbon and
nitrogen enrichment. Their hydrogen depletion varies between undetectable hydrogen to
severely depleted (5\% by mass). The
declines in brightness are caused by  the formation of
clouds of carbon dust in front of the star (see Clayton 1996 for a review).

To explain the existence of RCB stars, two scenarios have been proposed.
In the first scenario the RCB star is a merger of
two white dwarfs (WD; Iben, Tutukov \& Yungelson 1996; Saio \& Jeffery 2000,2002), 
while in the second scenario
a He-shell flash (thermal pulse) in a pre-WD inflates the
remaining small envelope, leading to a born-again AGB evolution. 
During this evolution various mixing and nuclear burning
processes can cause dramatic changes of the surface abundances (Iben
et al. 1983). However, no post-AGB stellar evolution model  can
account for the predominance of helium in the atmospheres of most RCB
stars (Herwig 1999,2000; Pandey et al. 2000).
Therefore, the WD-merger scenario might be better at explaining the RCB class.
Unfortunately, this theoretical scenario has not
been worked out in detail yet and skepticism is fueled, for example,
by the long merging times of WD in binary systems (Iben et al. 1996). 

Some other classes of hydrogen-deficient post-AGB stars show RCB-like light variations
due to dust obscuration.
The born-again scenario is known to produce
stars with the {\it spectral and variability characteristics of RCB stars}, which is why
it was at first postulated that all RCB stars might be created this way. In this class, 
we find Sakurai's Object (Duerbeck \& Benetti 1996),
V605 Aql (Clayton \& De Marco 1997) and FG Sge (Gonzalez et al. 1998), all of which were witnessed 
to undergo a born-again event and develop RCB spectral and light-curve characteristics. 
These stars, therefore, create a strong link between the RCB and the born-again classes.

A second class of stars, observationally linked to the RCB stars, is the 
Wolf-Rayet central stars of planetary nebula (PN)
\footnote{These are called [WR] stars after van der Hucht et al. (1981),
to distinguish them from massive
Wolf-Rayet stars, or [WC] stars to emphasize that they are Wolf-Rayet star of the carbon, rather than nitrogen
sequence.}. The link is established by two objects:
the [WC10] central star CPD--56$^{\rm o}$~8032 (De Marco, Barlow \& Storey 1997),
which has the same type of visual declines (albeit shallower: $\Delta V \sim$1~mag; Jones et al. 1999)
as RCB stars and the peculiar [WC12] (Crowther, De Marco \& Barlow 1998) V348 Sgr,
which is one of the most active RCB stars. The latter star, is also suspected to be
a born-again star because its effective temperature could not maintain the PN's
observed degree of ionization (Pollacco, Hill \& Tudhunter 1990).

The problem is therefore to understand what fraction of RCB stars
share a common evolution (WD-merger scenario or He-shell flash) 
and what fraction simply
shares the light-curve behavior and is grouped in the RCB evolutionary class for lack of a better explanation.
One might argue that the abundances should suffice as a discriminator. However 
beside the abundances, there are other characteristics which should be taken into
account when grouping these stars, because they seem to dictate different grouping rules.
In this paper, we concentrate on the above-mentioned V348 Sgr, which, together with
DY Cen, MV Sgr and HV~2671 (in the LMC; Alcock et al. 1996), 
constitute the class of hot RCB stars ($\sim$15\,000-20\,000~K). 
In both scenarios presently considered for the evolution of RCB stars (born-again 
or merging WDs), cool RCB stars 
should proceed at roughly constant luminosity toward the hotter
region of the HR diagram.
Therefore the hot RCB stars might be reasonably thought to 
represent the progeny of the cool ones.

In this paper, we consider the hot RCB stars in all their observable aspects, to try
and place them in relation to other post-AGB stars. 
The abundances of the hot RCB stars, suggest that two of them,
DY Cen and MV Sgr, should be classified as
successors of the typical He-dominated cool RCB stars while
V348 Sgr and HV~2671 might belong to the post-AGB born-again
group. However, additional characteristics 
do not unanimously support these sub-groups.
 
After describing new and archival observations
in Sec.~\ref{sec:observations}, we review the optical spectra of the four stars in Sec.~\ref{sec:spectra}.
In Sec.~\ref{sec:ObservedParameters}, we review the observed parameters of the stars, 
while in Sec.~\ref{sec:stell_par}, we summarize the
stellar parameters deduced from a mix of techniques.
In Sec.~\ref{sec:disk}, we interpret the split line profiles with equatorial enhancements.
In Sec.~\ref{sec:neb_spec}, we discuss the nebular properties.
In Sec.~\ref{sec:lightcurve}, we analyze the light-curve behavior. 
In Sec.~\ref{sec:SecularSpectralVariabilityOfTheHotRCBStars}, 
we discuss the spectral variability. 
We finally summarize our analysis in Sec.~\ref{sec:summary}, compare the observations with
the evolutionary models in Sec.\ref{sec:CommparisonWithEvolutionaryModels}, 
leaving some final remarks to Sec.~\ref{sec:TheEvolutionaryHistoryOfTheHotRCBStars}.

\section{Observations}
\label{sec:observations}

\subsection{New Data}
\label{ssec:NewData}

A spectrum of V348 Sgr was obtained on 1998, May 14 on the 2.5m Isaac Newton Telescope 
at the Observatorio Del Roque De Los Muchachos using the IDS spectrograph.  
The observation was taken with the 235 mm camera and the 
300 l/mm grating giving a resolution of 6~\AA. 
The spectrum has been extracted, 
flat-field corrected, sky subtracted, and wavelength and flux calibrated (with respect to the
standard star BD+26$^{\rm o}$~2606).

New spectroscopic observations were obtained of HV~2671
on 1997, December 1-2, when it was at light maximum.  
The spectra were obtained with the
SIT/CCD spectrograph on the SAAO 1.9m telescope at Sutherland, South 
Africa.
The grating used gives a reciprocal dispersion of 210 \AA~mm$^{-1}$ and a
resolution of approximately 5 \AA~with a 250 \micron~slit,
 giving a useful range of about
3500-7500~\AA\ at the angle setting used. The spectrum has been extracted,
flat-field corrected, sky-subtracted and wavelengthr- calibrated.
Flux calibration is done by observing the
standard star LTT~377. The flux calibration is not accurate because
the instrument is not a spectrophotometer. However since the observing conditions
were photometric, we believe that
the slope of the spectrum should be reliable.

Two blue spectra (3900-4800~\AA) of DY Cen were obtained at the Anglo-Australian
Telescope (AAT) on 2001, April 4 and September 9, with the RGO spectrograph and 25cm
camera. The 1200R grating was used in second order to give an effective
resolution for 2 pixels of $\sim$0.5\AA\/ on the 4096x2048 EEV42 detector. 
The data were corrected for pixel response and wavelength calibrated
using Argon lamp images taken before and after the stellar exposures.
After sky subtraction, the 1-dimensional stellar spectrum was extracted.

\subsection{Archival Data}
\label{ssec:ArchivalData}

The previously published and new visible light spectra for the sample stars are 
listed in Tables~\ref{tab:spectra1} and \ref{tab:spectra2}. 
We have examined in detail those of the published spectra that 
could be made available to us by the authors in electronic
format. 
The spectra on plates listed in Tables~\ref{tab:spectra1} and \ref{tab:spectra2} were located in the
Lick plate archive and then scanned into {\sc tiff} files with the kind help of
Tony Misch and George Herbig. A cut was then made through the {\sc tiff} files
and a rough wavelength calibration was applied. This method works well in
providing spectra that can be easily compared in a {\it qualitative} manner to
more modern spectra.                                                                                                                

In particular we made use of the following spectra (marked with an `x' in Tables~\ref{tab:spectra1} and \ref{tab:spectra2}):
For HV~2671 we used a 1995 spectrum (Alcock et al. 1996) as well as the new spectrum described
in Sec.\ref{ssec:NewData}.
For V348 Sgr we used a high resolution spectrum taken in 1987 
(Leuenhagen, Heber \& Jeffery 1994, which we will refer to as LHJ94), a lower 
resolution spectrum taken in 1995 (W.-R. Hamann, 
priv. comm.), as well as the low resolution spectrum taken by us in 1998 (Sec.~\ref{ssec:NewData}). 
For MV Sgr, we analyzed
a high resolution spectrum taken in 1995 (N.K. Rao, priv. comm.) and the two 1994 spectra of Venn et al. (1998).
For DY Cen,
we obtained two spectra, taken in 1989 and 1992, respectively (Rao Giridhar \& Lambert 1993, Giridhar Rao \& Lambert 1996).
All the analyzed spectra were taken when the stars were  at maximum light.

Low resolution UV spectra taken with IUE, exist
for V348 Sgr, MV Sgr and DY Cen (Heck et al. 1982; Evans et al. 1985; Drilling \& Sch\"onberner 1989; 
Goldsmith et al. 1990; 
Jeffery 1995; Drilling et al. 1997).
DY Cen has only one long wavelength IUE spectrum. In addition, HST FOS spectra exist
for V348 Sgr during a decline (two observations during a shallow minimum and one 
during a deep minimum; Hecht et al. 1998). We have re-analyzed the maximum-light IUE spectra
for the three stars.

\section{Optical and UV Spectra}
\label{sec:spectra}

In Figs.~\ref{fig:atlas1} and \ref{fig:atlas2},
we present the optical spectroscopy of the three Galactic
hot RCB stars, while in Fig.~\ref{fig:spec1} we show the spectrum of HV~2671 compared to our 
low resolution spectrum of V348 Sgr.
Figs.~\ref{fig:atlas1} and \ref{fig:atlas2}
contain a mix of the spectra available, so as to obtain the largest spectral range in common to at 
least two stars. In particular, for MV Sgr we present the high resolution 1995 spectrum.
In the spectral range 3800-4200~\AA\ this spectrum is compared to the
intermediate resolution Calar Alto spectrum of V348 Sgr, while in the range 4200-5000~\AA , 
it is compared with the high resolution
LHJ94 spectrum of V348 Sgr. The DY Cen spectrum in Fig.~\ref{fig:atlas1} is from April 2001.
In the range 5600-6000~\AA , we compare MV Sgr to the lower resolution Calar Alto
spectrum of V348 Sgr (no high resolution spectrum of V348 Sgr exists in this spectral range) and the 1992 spectrum of DY Cen. 
Finally, in the range 6000-6800~\AA , we compare MV Sgr to the high resolution V348 Sgr spectrum and the 1992
spectrum of DY Cen. All stars have been shifted according to the measured radial velocities 
(Sec.~\ref{sec:ObservedParameters}).
Line identifications
have been taken from Leuenhagen et al. (1994), Giridhar et al. (1996) and Jeffery et al. (1988). 
The forbidden line spectra are discussed in Sec.\ref{sec:neb_spec}.

\subsection{V348 Sgr \& HV~2671}

The visible maximum-light spectrum of V348 Sgr resembles a very cool [WC] 
central star of PN, although Crowther et al. (1998) have argued that this object simply fails the 
[WC] classification in that it has no C~{\sc iii} $\lambda$5696. They have proposed 
a peculiar extreme helium star classification. However, the overall spectrum of V348~Sgr
at maximum light is somewhat similar to (although cooler than) the weak-lined [WC11] central star
K2-16 (Crowther et al. 1998). It is also interesting that although many of the [WC] central
stars have {\it not} been checked for variability, of those that have been monitored, only
CPD-56\arcdeg~8032 (a [WC10] star) has shown small RCB-like declines of about 1 
magnitude (Pollacco et al. 1992; Jones et al. 1999).

The strongest emission lines in the spectrum of V348 Sgr are from C~{\sc ii} (pure emission,
no P-Cygni profiles) and He~{\sc i} (all with P-Cygni profiles). 
Lines of C~{\sc iii} are absent as are those of He~{\sc ii}, while some lines of C~{\sc i} 
are present.
All the Balmer lines in our spectral range are present in emission with P-Cygni profiles. 

The spectra of HV~2671 and V348~Sgr 
are, at 5~\AA\ resolution, almost identical (Fig.~\ref{fig:spec1}). The only exception is that the
carbon emission spectrum looks stronger relative to the hydrogen and helium lines in V348 Sgr. 
This could be indicative of a larger carbon abundance in this star. No obvious nebular lines
are detected in the spectrum of HV~2671 (see also Sec.~\ref{sec:neb_spec}).

\subsection{DY Cen \& MV Sgr}

Giridhar et al. (1996) discuss their two spectra of DY Cen pointing out several remarkable
differences. While the differences between the two spectra
are discussed in depth in Sec.~\ref{sec:SecularSpectralVariabilityOfTheHotRCBStars}, 
there are some general characteristics which remain unchanged.
The spectrum of DY Cen is dominated by simple absorption lines of C~{\sc ii},
N~{\sc ii}, Ne~{\sc i} and Al~{\sc ii} and {\sc iii}. At the same time, some C~{\sc ii} lines (e.g. M2 and M5)
and the He~{\sc i} lines, have
a split absorption (or an absorption with a reversed center) in 1989, while in 1992 they appear 
as weak split emission lines with weak
P-Cygni absorptions. H$\alpha$ is in emission with a P-Cygni absorption. 

Although the spectrum of DY Cen often exhibits absorption lines where the spectrum of V348 Sgr has emission lines, the two
spectra are much closer in appearance because more often than not they have 
lines in common. On the other hand, although the spectrum of
MV Sgr exhibits a wealth of emission lines, the strongest of them almost never coincide with strong lines
in V348 Sgr, giving MV Sgr a completely different appearance to the other two stars. 
The strongest emission lines in the spectrum of MV Sgr belong, in
fact, to metals (the lines of Fe~{\sc i} and
{\sc ii}, Si~{\sc ii}, Ni~{\sc i}, Ca~{\sc i} and Li~{\sc i}, are all split and have a peak separation of $\sim$100~km~s$^{-1}$,
Pandey et al. 1996), with carbon lines in absorption,
and helium lines in weak emission. H$\alpha$ is strongly in emission.

\vspace{1cm}

Visible spectroscopy of RCB stars when they are below maximum light is rare.
Herbig (1958) found that as V348 Sgr faded, the C~{\sc ii} emission also faded to be replaced 
by a nebular spectrum including
strong Balmer lines and [O~{\sc ii}] $\lambda$3727.  By the time V348 Sgr was $m_V \sim 17$~mag, the 
C~{\sc ii} lines had disappeared.   
Dahari \& Osterbrock (1984) were able to see the C~{\sc ii} lines at 
minimum light ($m_V$ = 18.4~mag), but
they were very weak compared to maximum-light, and weaker than the nebular lines. 
No spectra have been obtained of the other three stars when they were below maximum light.

The low resolution, long wavelength IUE spectra of V348 Sgr, DY Cen and MV Sgr at maximum light
are all very similar, once brightness and reddening are taken into account (Sec.~\ref{sec:ObservedParameters}). 
The most prominent line in that range is an emission line
of Mg~{\sc ii} at 2800~\AA . The low resolution, short wavelength, maximum-light 
spectra of MV Sgr and V348 Sgr are also quite similar 
(there is no short wavelength spectrum for DY Cen). The most prominent lines appear to be
absorptions of C~{\sc ii} $\lambda$1334.5 ($^2$P-$^2$D) and C~{\sc iv} $\lambda$1550 ($^2$S-$^2$P$^o$).
Hecht et al. (1998) have published HST FOS spectra of V~348 Sgr, during shallow and deep minimum light.

\section{Observed parameters}
\label{sec:ObservedParameters}

In this Section, we discuss the observed parameters of the hot RCB stars. These include the 
radial velocity of the star, the observed $V$ brightness and the reddening. 
For the LMC hot RCB star HV~2671 we also include the distance 
(which is accurately determined) and the resulting 
$M_V$ value, among the observed quantities. For the other three stars, the distances and
$M_V$ values are discussed,
together with the stellar models, in Sec.~\ref{sec:stell_par}. Although in the text we discuss most of the
literature  parameters along with our own-derived values, in Table~\ref{tab:basic} we list
only the adopted values.

\subsection{V348 Sgr \& HV~2671}

From the LHJ94 spectrum, we determine an LSR (heliocentric) radial velocity for V348 Sgr of 
142~km~s$^{-1}$ (130~km~s$^{-1}$) from pure
emission lines (the O~{\sc ii} M24 component at 4751.3~\AA , the C~{\sc ii} M22 component at 5836.4~\AA, and the only un-blended 
nebular line, [N~{\sc ii}] $\lambda$6543). When carrying out the same measurement on lines with obvious P-Cygni
absorptions (e.g., H$\alpha$), the resulting LSR (heliocentric) radial velocity is 157~km~s$^{-1}$ (145~km~s$^{-1}$). 
This is due to the fact
the P-Cygni absorption eats into the blue side of the emission lines, displacing their apparent position to the red. 
There are many estimates of the reddening of V348 Sgr (see Table~\ref{tab:red}). 
Of these, the Balmer decrement estimates of Dahari \& Osterbrock (1984), and Houziaux (1968) are
suspect because the nebular lines are contaminated by the stellar hydrogen emission. Pollacco et al.
(1990) observed the nebula off the star and could therefore determine more reliable nebular fluxes. 
Most of the other estimates lie within 0.2~mag
of the Pollacco et al. (1990) estimate. We adopt their value of $E(B-V)$ = (0.45 $\pm$ 0.1)~mag.
This value nulls effectively the 2200-\AA\ feature in the low resolution IUE spectrum of this star.
At present, the maximum light magnitude of V348 Sgr is 
$m_V$ = 12.0~mag (American Association of Variable Star Observers [AAVSO]). 

Alcock et al. (2001) estimate the
reddening toward the bar of the LMC to be $E(B-V)$ $\sim$ 0.17~mag. This would also be the reddening 
toward HV~2671, in the absence of circumstellar reddening. The reddening 
toward HV~2671 was also determined from intrinsic colors to be $E(B-V)$ $\sim$ 0.3~mag (Alcock et al. 1996). 
Alternatively, we can determine the reddening to HV~2671 by comparing the slope of
its flux-calibrated spectrum,
with that of V348 Sgr, de-reddened by our adopted $E(B-V)$ = 0.45~mag. An almost perfect match is
derived for a value of $E(B-V)$ = 0.15~mag. 
Given that the rectified spectra are extremely similar (Fig.~\ref{fig:spec1}), this is thought
to be a reasonable method.
The two estimates agree within the uncertainties involved. We adopt the latter value for this study.
This implies that there is a minimum amount of circumstellar reddening.
The brightness of HV~2671 at maximum light is $m_V$ = 16.1~mag (Alcock et al. 2001). 
Then, using the LMC distance modulus ($m-M$ = [18.6$\pm$0.1]~mag; Feast 1999) and assuming $R_V$ = 3.1, 
the absolute $V$ magnitude of HV~2671 is therefore $M_V$ = --3.0~mag. 

\subsection{DY Cen \& MV Sgr}

From the absorption line spectrum of DY Cen in 1989, Giridhar et al. (1996) determined the radial velocity
to be (41$\pm$4)~km~s$^{-1}$ from 25 C~{\sc ii}, 14 N~{\sc ii} and 20 Ne~{\sc i} lines. Similar measurements
on the 1992 spectrum returned a value of (29$\pm$3)~km~s$^{-1}$. 
From the nebular lines [N~{\sc ii}] $\lambda$$\lambda$6548,83, 
[O~{\sc i}] $\lambda$6300, [S~{\sc ii}] $\lambda$$\lambda$6717,31 and H$\alpha$, they determined mean values
of (23$\pm$2)~km~s$^{-1}$ and (22$\pm$2)~km~s$^{-1}$ for 1989 and 1992, respectively.
From the emission components of 8 C~{\sc ii} lines, they determined 
radial velocity values in the range 28-123~km~s$^{-1}$ in 1989 and 20-32~km~s$^{-1}$ in 1992. Giridhar et al. (1996)
also used He~{\sc i} lines which, in 1989, were all in absorption, returning a value for the radial velocity 
in the range 4-53~km~s$^{-1}$, while in 1992 were in emission with P-Cygni absorption. From the emission part
they determined radial velocities in the range 2-55~km~s$^{-1}$. 

We adopt the radial velocity value determined from the nebular emission lines, the
least affected by stellar variability, as the most representative of the 
systemic velocity of the star. We should also note that the absorption lines in each of the two epochs, 
have very little radial velocity scatter, but there is a significant change between 1989 and 1992. 
This could be understood if DY Cen is a spectroscopic binary
(as also suggested by Giridhar et al. [1996]). If this fact could be corroborated, {\it it would be the first confirmation
of binarity in an RCB star}. The fact that the emission lines (or lines that developed emission in 1992)
have such a large radial velocity scatter will be amply discussed in Sec.~\ref{sec:SecularSpectralVariabilityOfTheHotRCBStars}.
The reddening was determined by Jeffery \& Heber (1993) to be $E(B-V)$ = 0.50~mag, from a stellar continuum analysis.
When this value is used the slope of the IUE spectrum of DY Cen aligns with that of the de-reddened 
spectrum of V348 Sgr. The maximum-light brightness of DY Cen is $m_V$ = 13.0~mag (AAVSO).

The radial velocity of MV Sgr was measured from the 1995 spectrum. From the absorption lines
of C~{\sc ii} $\lambda$4267 and He~{\sc i} $\lambda$4437, we obtained un-corrected values
of --72.5 and --60.7~km~s$^{-1}$, respectively. From the split emission line of Fe~{\sc ii}
$\lambda$5991, we obtain --62.2~km~s$^{-1}$, while from the strong emission line of Ca~{\sc ii}
at 8542~\AA\ we obtain 66.7~km~s$^{-1}$. 
A mean of these values, corrected for LSR
(heliocentric) radial velocity correspond to (--83$\pm$5)~km~s$^{-1}$ ([--95$\pm$5]~km~s$^{-1}$).
This compares favorably with the value of (--93$\pm$4)~km~s$^{-1}$,
determined by Pandey et al. (1996) from their 1992 spectrum (we presume their value to be corrected for
the heliocentric standard of rest, although they never state that explicitly). This also agrees well with
earlier estimates of Jeffery et al. (1988; [--91$\pm$7]~km~s$^{-1}$ from the absorption lines, 
[--81$\pm$17]~km~$^{-1}$ from the emission lines) and Rao, Giridhar \& Nandy (1990; [--95$\pm$8]~km~s$^{-1}$).
The reddening toward MV Sgr was estimated by Heber \& Sch\"onberner (1981) from a comparison of observed
(Landolt 1979) and predicted colors, to be $E(B-V)$ = 0.45~mag. Drilling et al. (1984) determine
$E(B-V)$ = 0.38~mag from the strength of the 2200-\AA\ bump. From comparing the de-reddened ($E(B-V)$ = 0.45~mag)
UV spectrum of V348 Sgr to that of MV Sgr we determine $E(B-V)$ = 0.43~mag. The maximum-light brightness of MV Sgr
is $m_V$ = 13.4~mag (AAVSO).

\section{Stellar Model Parameters}
\label{sec:stell_par}

In this Section, we summarize the stellar parameters of the hot RCB stars from the literature. For
all the hot RCB stars except for HV~2671, these have been determined from modeling of the stellar continuum and
lines. For HV~2671, we derive the parameters by analogy to those of V348 Sgr, but accounting for the known distance to
the LMC. For this star we have the only truly reliable estimate of $M_V$. The selected parameters are 
summarized in Table~\ref{tab:model}.

\subsection{V348 Sgr \& HV~2671}
\label{ssec:stellar_parameters_v348}

Leuenhagen \& Hamann 
(1994) modeled the UV and optical spectra of V348 Sgr with their non-local thermodynamic equilibrium (LTE) code for WR
atmospheres (Hamann \& Koesterke 1993). The stellar parameters they derive are summarized in
Table~\ref{tab:model} for their adopted distance of 5.4~kpc (a value which cannot presently be improved
upon). Beside what is listed in the table, they find $v_{\infty}$ = 190~km s$^{-1}$ 
using a beta velocity law with $\beta$ = 2, and
a neon abundance of 2\% (by mass). Their hydrogen abundance of 4\% (by mass), might be an upper limit.
The H$\alpha$ line, on which they base their abundance estimate, is in fact a
blend of stellar and nebular emission. Although the nebular line
spectrum during maximum light is faint, we can estimate the nebular
contribution to the observed H$\alpha$ emission line in the following way.  We measured
the equivalent width of the [N~{\sc ii}] line at 6548~\AA , which can
easily be de-blended from the neighboring stellar N~{\sc ii} M45
component with a double Gaussian. This is (0.25$\pm$0.05)~\AA .  From
Pollacco et al. (1990) we see that the nebular H$\alpha$ in their
spectrum is 6.2 times stronger than the [N~{\sc ii}] $\lambda$6548
line.  From this we can estimate the equivalent width of the nebular
H$\alpha$ contribution to be (1.55$\pm$0.40)~\AA . The total
equivalent width of the H$\alpha$ feature in our high-resolution
spectrum is (2.25$\pm$0.22)~\AA , so that the stellar contribution to
this feature is predicted to be (0.68$\pm$0.46)~\AA , or less than
half of the total strength.  The hydrogen abundance derived by
Leuenhagen \& Hamann (1994) is therefore a comfortable upper limit and
so, we note that this star might be much more hydrogen-poor than
previously thought. Code improvements (such as the addition of line
blanketing) could easily lead to slightly different parameters, although
there is no reason to believe that the stellar picture would 
be fundamentally changed. A more serious unknown is the departure from
spherical symmetry unaccounted for by the models (Sec.~\ref{sec:disk}) and its effects on the line intensities
used to determine the stellar parameters.

V348 Sgr contains absorptions as well as emission lines. Therefore,
plane-parallel codes can also be used to determine its stellar parameters and abundances.
This provides us with a formidable check that fundamentally different codes such
as spherical, expanding transfer codes and plane parallel static codes
do indeed return parameters in the same ball park. Without this check one might
fear that the different abundances obtained for RCB stars and for [WC] stars
might in the end be an artifact.
From an LTE analysis of the absorption line spectrum,
Jeffery (1995) derived $T_{eff}$ = 22,000 K, 
log g$<$ 2.8, and assuming the hydrogen mass fraction of Leuenhagen et al. (1994), he
derived the following mass fractions:
$X_{He}$ = 0.67, $X_C$ = 0.28, with log$X_N$ = --2.73, log$X_O$ = --2.92 and log$X_{Si}$ = --4.27.
Although one could argue that the agreement is poor, we can none the less conclude that indeed
the abundances of V348 Sgr are closer to those of [WC] central stars than to the those of RCB stars
including DY Cen and MV Sgr (see later).

The temperature of HV~2671 has not been measured independently, but because of its similarity to V348 Sgr,
we adopt the same temperature (20\,000~K). Based on it and the related bolometric correction
of --1.71~mag, together with the observed $V$ magnitude at maximum light, the 
reddening of $E(B-V)$ = 0.15~mag determined in Sec.~\ref{sec:ObservedParameters} 
and the known distance to the LMC (50~kpc; Feast 1999), we determine
a bolometric luminosity for HV~2671 of $\sim$6000~L$_\odot$. 
The radius then follows from the Boltzmann formula. 
Based again on the similarity of the spectrum
at low resolution we have no indication that the abundances are not
similar to those of V348 Sgr. 

\subsection{DY Cen \& MV Sgr}

DY Cen was analyzed by Jeffrey \& Heber (1993). Through an LTE analysis of the absorption lines they derived 
effective temperature, gravity and abundances (listed in Table~\ref{tab:model}). 
The abundances of DY Cen are more similar to the mean abundances of other RCB stars than to V348 Sgr.
They also derive a mass
of 0.8~M$_\odot$ assuming that the stellar interior consists of a CO core. This value of the
mass, together with their derived value of the gravity implies a radius of the star of 12.5~R$_\odot$, and a 
luminosity of 20\,200~L$_\odot$. They also establish an upper limit for $v \sin i$ of 20~km~s$^{-1}$.
The emission line spectrum of DY Cen was examined by Giridhar et al. (1996), who reported several variations between
those two epochs (Sec.~\ref{sec:SecularSpectralVariabilityOfTheHotRCBStars}). 
No analysis has ever been carried out of these emission lines.
As for the distance to this object, Rao et al. (1993) quote a range of 1.4-4.8~kpc. At 4.8~kpc, with our own values of
the observed $V$ magnitude and preferred reddening ($E(B-V)$ = 0.5~mag) we determine $M_V$ = --2.0~mag. This is much
fainter than for the other three stars. We therefore suggest that the distance to DY Cen might be only a lower limit.

The stellar parameters of MV Sgr have been derived
by Drilling et al. (1984) and Jeffery et al. (1988) from plane-parallel
LTE analyzes. 
In the analysis of Drilling et al. (1984) we find an estimate of the
temperature based on a fit to UV 
spectrophotometry and UBV photometry. They derive T$_{\rm eff}$ =
(16\,000$\pm$500)~K and $E(B-V)$ = 0.38~mag for
a de-reddened $V$ magnitude of 11.9, derived from a flux of 6.4 x
10$^{-14}$ erg cm$^{-2}~s^{-1}~\AA^{-1}$. Assuming $E(B-V)$ = 0.38~mag, this
corresponds to an
observed maximum-light $V$ magnitude
of 13.1~mag. This observed value is not from Glass (1978), as cited in
Drilling
et al. (1984).  Apparently, they used an observed value of $V$ = 12.70~mag from
Herbig (1964).
The discrepancy between these values cannot be explained. 
Using  this temperature Jeffery et al. (1988) derive the abundances from
fits to the 
absorption lines (Table~\ref{tab:model}). They also
derive a surface gravity log(g) = 2.5$\pm$0.3 by analogy with HD 124448,
admitting to a substantial uncertainty.
The hydrogen abundance is quoted as being small. Herbig (1964) says there
is no sign of hydrogen absorption.
Pollacco \& Hill (1990) state that there is no doubt about the hydrogen
deficiency.

The emission lines of MV Sgr were studied by Pandey et al. (1996),
although no parameters are derived from them. Heber \& Sch\"onberner (1981) derive a temperature of 19\,000~K,
using a reddening $E(B-V)$ = 0.46~mag, from which they derive $M_V$ = --3.9~mag and a distance of 7.8~kpc. We have not 
added these
values to the table since the abundance analysis is dependent on the temperature and reddening values of
Drilling et al. (1984), while the other estimates (like the distance)
are inter-related with the 
higher temperature determined
by Heber \& Sch\"onberner (1981).

\section{The split lines in the spectra of V348 Sgr, DY Cen and MV Sgr: evidence for an equatorial buldge}
\label{sec:disk}

It was noticeable that the double peaked line profiles seen in  the  C~{\sc ii} 
and He~{\sc i} lines of V348 Sgr \& DY~Cen and the C~{\sc i}, 
He~{\sc i} and H~{\sc i} lines  of MV~Sgr (Pandey et al. 1996) closely resemble 
those observed in  some Be stars (where they are assumed to 
arise in an equatorial disk), and therefore suggest  a departure from  spherical  symmetry for the mass out-flows from these 
stars. However,  despite the similarities of the emission line profiles   of Be stars  presented by Dachs, Hummel \& 
Hanuschik (1992) to those presented here we do not believe  the circumstellar  environment of the  hot RCB stars are exactly  
analogous  (although we do suspect an equatorial density  enhancement for the hot RCB stars). 

Current theories for the behavior of Be star disks suggest that they are geometrically thin quasi Keplerian excretion disks
 that are supported by the input of angular momentum and material at the disk/star boundary (e.g. Porter 1999). However, the 
line profiles observed for MV~Sgr and V348~Sgr differ in three respects from the predictions of a {\em Keplerian} disk. 

\begin{enumerate}
\item
For a Keplerian disk one would expect the weaker emission lines to be produced in a smaller region of the  
disk than the stronger lines and hence have a larger line width due to the greater rotational velocities of the inner regions of 
such a disk  (e.g. the weaker Fe~{\sc ii} lines have a greater width than  H$\alpha$ for Be stars disks; Dachs, Hummel \& 
Hanuschik 1992). However  the line widths of the weaker emission lines are systematically lower than those of stronger 
transitions for MV~Sgr (Pandey et al. 1996).

\item
The high resolution line profiles presented for MV~Sgr by Pandey et al. (1996) show large asymmetries in the 
peak ratios of the double peaked line profiles. In classical Be stars this is attributed to the slow precession of a one armed 
density wave within the Keplerian disk (e.g. Okazaki 1997) and the peak ratios of different lines  are observed to vary 
in phase (such that for a given time the blue peak is stronger than the red [or vice versa] for all lines). The line profiles for
 MV~Sgr clearly show that this is not the case, and the asymmetries cannot arise from the precession of a density wave within a 
Keplerian disk.

\item
The P-Cygni components of the line profiles of MV~Sgr and V348~Sgr are {\em not} observed for classical Be stars. Indeed, there is
strong evidence that the radial component of the disk velocity is highly subsonic ($<$1~km~s$^{-1}$) until large radii 
(of the order of 100R$_{\ast}$; Okazaki 2001), well outside the region responsible for line formation. Consequently
the presence of the P-Cygni absorptions at $\sim$100~km~s$^{-1}$ in  MV~Sgr and V348~Sgr cannot arise in a disk of similar 
structure to those of classical Be stars. 
\item
Neither active dust nor forbidden line emission is observed from the  equatorial {\em Keplerian excretion} disks  found around 
classical Be stars.

\end{enumerate}

However, despite  the differences between classical Be stars and hot RCB stars the  presence of an equatorial density enhancement 
of some form is attractive for a number of reasons. Polarimetric observations of R CrB suggest the presence of an obscuring 
dusty torus (Rao \& Lambert 1993, Clayton et al. 1997), while changes in the polarization of other RCB stars during deep declines suggest the 
observed light is scattered (e.g. Clayton 1996, and references within) - obscuration of the star by an equatorial torus with 
scattering of light from the polar axis into the line of sight would explain these observations. 
The formation of dust in a spherical wind is also difficult  because at large radii, where the temperature is low enough to
 allow dust to condense ($\sim$3000~K) for mass-loss rates as low as those observed for HV~2671 \& V348~Sgr (Table~\ref{tab:model}), the wind 
density is simply too low to allow dust formation (e.g. Bjorkman 1998); a `disk' or `torus' could permit high enough densities 
to exist for dust to form. Additionally, the higher densities allow more efficient cooling - modeling of the  energy balance of 
geometrically thin disks around Be stars by Millar \& Marlborough
(1999) suggest that 
they are approximately isothermal with a 
temperature of $\sim$0.5T$\ast$. With additional cooling due to the high metal content of hot-RCB winds and the possibility of 
self shielding produced by a geometrically thick disk the temperature within such a structure may be low  enough to permit 
dust formation.

A class of objects where the same problem of dust production in an hostile environment occurs are the B[e] stars, 
stars of spectral type B with a rich emission line spectrum and  pronounced IR excess due to the presence of equatorially 
concentrated hot dust (the presence of forbidden line emission and hot dust distinguishes these objects from  classical  
Be stars). However Galactic B[e] stars are a rather heterogeneous class due to  difficulties in determining absolute luminosities 
and comprise both pre- and post-Main Sequence stars, with luminosities ranging from log($L_{\ast}/L_{\odot}$) = 2.5--6.5 
(Zorec 1998). Lamers et al. (1998) provide a summary of known objects and conclude that 12 Galactic B[e] stars are likely to be 
compact PNe, while a smaller number appears to be post-AGB objects - Hen~3-1191 (Lamers et al. 1998), 
Hen3-847 (Miroshnichenko priv. comm.), MWC~922 and MWC~300 (Voors 2000). Given that the hot-RCB 
stars share many of the same  observational properties as the B[e] stars (indeed MV~Sgr meets the  membership criteria listed by 
Lamers  et al. [1998]) and other objects initially classified as B[e] stars are now recognized to be post-AGB objects, 
it is tempting to suggest  that the hot-RCB stars share a similar circumstellar environment as the B[e] stars.

Inspection  of the spectrum  of the B[e] star HD~87643 (Oudmaijer et al. 1998) 
shows that the profiles of the H~{\sc i} Balmer series 
closely resemble those of the strong emission lines of C~{\sc i}, He~{\sc i} and H~{\sc i} 
in MV~Sgr which also show 2 absorption components; a 
pronounced P-Cygni absorption trough (at $\sim$100km~s$^{-1}$ - the larger velocity of the blue edge of the P-Cygni profile in 
HD~87643 reflecting a higher terminal  velocity) and the central reversal (which  appears to be slightly  blue shifted 
in both HD~87643 and  MV~Sgr; Pandy et al. 1996). Additionally, 
as with MV~Sgr, the weaker Fe~{\sc ii} lines in HD~87643 
are observed to be half the width of the  Balmer lines, while the observable forbidden lines are narrower still.
The same is also true of V348~Sgr and DY~Cen although we note that the central reversals in these stars do not show any 
velocity  shift, apparently indicating absorption by $\sim$stationary material. We note that  this behavior is also observed 
for the Be shell stars which have deep central absorption troughs  - it is thought that these stars 
represent the subset of classical Be stars that are seen at high inclinations where the line of sight lies close to or in  the 
equatorial plane  and hence intersecting the circumstellar disk.

Oudmaijer et al. (1998) interpret this  behavior  as a result of  the presence of at least 2 and probably 3 different wind 
regimes; a very strong high  velocity wind  responsible for the broad Balmer lines  (and the P-Cygni absorption), a second,
lower velocity wind traced by the Fe~{\sc ii}  lines and the core of the Balmer lines (the central absorption is  attributed to self 
absorption in a slow wind rather than Doppler  splitting due to rotational motion).  The final component, sampled by  the narrow 
forbidden lines is attributed to a possible low  density nebula (c.f. V348~Sgr and DY~Cen). Furthermore, changes in the  
polarization across the H$\alpha$ line and adjacent continuum  clearly indicates a departure from spherical  symmetry and
indicate the  presence of both expansion and rotational velocity components  (note that line emission from the classical Be  
star disks shows {\em} no  evidence for radial expansion). Oudmaijer et al. (1998) suggest that the varying wind regimes can 
be accommodated in a scenario where a radiation driven wind from a slowly expanding dense  circumstellar disk (responsible 
for  the Fe~{\sc ii} lines and central  absorption in the Balmer lines) interacts with a faster polar wind  (responsible for the 
P-Cygni absorption line components) -  given the similarities in the line profiles of the hot-RCB stars to those  of 
HD~87643 we suggest a similar hybrid  wind regime could be present in  these stars as well. 

The pronounced double peaked Fe~{\sc ii} profiles and flat topped forbidden lines observed in  MV~Sgr by Pandy et al. (1996) 
are not  found for HD~87643 by Oudmaijer et al. (1998); however we suggest that the permitted Fe~{\sc ii} lines may arise from a 
disk wind while the narrower forbidden lines may arise at relatively large radii where the distinction between the 
different regimes has largely disappeared. Although this seems somewhat contrived we note that flat topped optically thin line
profiles typically arise in  quasi-spherical outflows - therefore whatever mechanism is invoked to explain the 
double peaked Fe~{\sc ii}, He~{\sc i} and C~{\sc i} lines must also allow for a quasi spherical outflow, which  necessitates a change in 
geometry at large radii (since the forbidden lines must arise in a  low density region of the outflow). While the very narrow 
forbidden line emission in HD~87643 is absent from MV~Sgr, we note that it is  present in both DY~Cen and V348~Sgr suggesting the 
presence of a slow moving `nebular' component to the circumstellar environment of all three systems.

In Fig.~\ref{fig:pedestal} we compare all the V348 Sgr C~{\sc ii} lines which
show the split (namely $\lambda$4267.3 $\lambda$$\lambda$6578.1,6582.9 and $\lambda$$\lambda$5889.78,5891.6),
as well as He~{\sc i} $\lambda$5876. From this figure it can be seen that the components are all at similar radial 
velocity, separated by about (70$\pm$5)~km~s$^{-1}$. 
We have fit C~{\sc ii} $\lambda$4267 and He~{\sc i} $\lambda$5876 (the strongest, least blended
lines in the sample) with a combination of Gaussians, to aid in their interpretation. 
The central wavelengths of the fit components and their widths were free to vary.

We started by fitting the C~{\sc ii} $\lambda$4267 profile with three Gaussians, one for the broad pedestal and
one each for the two emission peaks. The fit is shown in Fig.~\ref{fig:hei5876} (a) (top) while the fit
parameters are listed on the left-hand-side of Table~\ref{tab:fit} (Test 1). Second (Test~2), we could interpret the 
line morphology {\it not} as a split line but as a {\it stellar} wind emission line with a broad base and a narrow 
top, blended with an absorption line from material intersecting the line of sight. We therefore fit the profile
(Fig.~\ref{fig:hei5876} (b)) with two emission Gaussians, a broad one for the base  and a narrower one for the rest of 
the line, after removing the central trough by snipping out the piece of spectrum. In this way, only the line wings 
were fit. We then subtracted the full observed profile from the fit obtaining the shape of the absorption line 
(Fig.~\ref{fig:hei5876} (c)). The resulting  absorption line was then fit with a cloud model
whose parameters we list in Table~\ref{tab:cloud}. From this exercise we concluded that if indeed there
is a shell of C~{\sc ii} which is absorbing flux from the emission line, it must be approximately stationary 
(the center of the line is at --3~km~s$^{-1}$) with an internal velocity dispersion of 20~km~s$^{-1}$ and a total 
column density of log(N/cm$^{-3}$) = 12.2 (Table~\ref{tab:cloud}). 

The same steps were carried out for the He~{\sc i} line at 5876~\AA . The only difference is that this line does not 
exhibit a broad pedestal. It does however have a significant P-Cygni absorption profile on its blue side. In
Table~\ref{tab:fit}, we list the fit parameters for the He~{\sc i} line
alongside those for the C~{\sc ii} fits.
The fit components which are directly comparable between the two lines are highlighted in bold. 
Line profile 1, in each of the two tests, represents the broad pedestal in the C~{\sc ii} case and the P-Cygni absorption
in the case of the He~{\sc i} line, and is therefore not comparable between the C~{\sc ii} and He~{\sc i} fits. 
We compare the two emission component positions and widths from fit Test 1. As we can see from Figs.~\ref{fig:hei5876} (a), 
the fits are excellent in both cases although the parameters vary considerably between the C~{\sc ii} and He~{\sc i} fits.  
The positions of components 2 and 3 in 
Test 1 are different in the C~{\sc ii} and He~{\sc i} lines (--43 and 37~km~s$^{-1}$ for C~{\sc ii} vs --22 and 
43~km~s$^{-1}$ for He~{\sc i}). 
The split lines' components do not reside at similar radial velocities
which excludes a Keplerian disk as their origin.
On the other hand, if we turn to fit Test 2, we see that
the position of the emission component is similar to that of the absorption trough.
Additionally, the number ratio of the column densities for the absorbing shells is 
C/He = 0.79, reasonably close to the value of 0.45 determined by Leuenhagen \& Hamann (1994) for the star. 
This interpretation is more in line with the equatorial density
enhancement more typical of B[e] stars.

Finally, we would expect the H$\alpha$ profile in the spectrum of V348 Sgr 
to also exhibit a double peak. The reason it does not is likely to be due to the fact
that the nebular H$\alpha$ is filling in the trough. Comparing the H$\alpha$ profile to
the C~{\sc ii} M1 component at 4745~\AA\ (Fig.~\ref{fig:ha_comp}) we see that the line widths are similar,
although H$\alpha$ is asymmetric, with more emission in the red part than the blue, somewhat more similar to
He~{\sc i} $\lambda$5876 (except for the split) than to the C~{\sc ii} profile. Additionally,
the P-Cygni profiles of the H$\alpha$ and He~{\sc i} lines align, showing their origin to be similar. 
In order to estimate the nebular contribution to the H$\alpha$ profile, we fit the 
[N~{\sc ii}] line at 6548~\AA\ to determine its equivalent width. We then multiplied 
the profile fit by 6.2, as explained in Sec.~\ref{sec:stell_par} and used it as a representation of the nebular 
contribution to the H$\alpha$ profile. The resulting predicted profile (H$\alpha_{\rm neb}^{\rm pre}$) 
is plotted in Fig.~\ref{fig:ha_comp} (b), together with the observed H$\alpha$ profile (H$\alpha_{\rm tot}^{\rm obs}$). 
We then subtracted the predicted nebular profile  from the total observed profile, obtaining the stellar contribution. 
The resulting profile (H$\alpha_{\rm tot}^{\rm obs}$ - H$\alpha_{\rm neb}^{\rm pre}$) is
also plotted in Fig.~\ref{fig:ha_comp}, displaced by --1.2 units, and compared to the He~{\sc i} $\lambda$5876
profile. 
Although the match is not perfect the peaks and trough do roughly
correspond.
This hints at the fact that the stellar component of the H$\alpha$ profile might be weak, split like the He~{\sc i} and
C~{\sc ii} components but almost completely overwhelmed by the nebular component of the same line.
We feel that further considerations would over-interpret the data.


\section{The Nebular Spectrum}
\label{sec:neb_spec}

In this Section, we review what is known about the nebulae around hot RCB stars. Unfortunately
the nebular spectra at maximum light is extremely weak, so that it is impossible to 
carry out a full nebular analysis. Additionally, unless spatial information can be derived, it is unclear
whether the forbidden lines arise far from the star, in a PN-type nebula, or whether they are formed in the
outer parts of a stellar wind.

\subsection{V348 Sgr \& HV~2671}

The first striking characteristic that distinguishes V348 Sgr, is that it has a large 
(30 arcsec, Pollacco et al. 1990), ionized nebula, much like an old PN. 
Pollacco et al. (1990) carried out a nebular analysis of the ionized gas around V348 Sgr,
by observing the nebula off the star, when the star was at minimum light. 
They observe lines of
[O~{\sc i}], [O~{\sc ii}] and [O~{\sc iii}] (although [O~{\sc i}] is flagged as of possible 
sky origin and [O~{\sc iii}] is extremely weak), [N~{\sc i}] and [N~{\sc ii}] 
([N~{\sc i}] as weak as [O~{\sc iii}]), [S~{\sc ii}] and the
Balmer lines. Additionally, they warn that the He~{\sc i} lines appear clearly extended in their spectrum, and
may therefore also be of nebular origin. 

The only nebular lines that could be measured in the LHJ94 spectrum of V348 Sgr
are [N~{\sc ii}] $\lambda$6548 and [O~{\sc i}] $\lambda$$\lambda$6300,6363. 
Their FWHM are 52~km~s$^{-1}$ and 19~km~s$^{-1}$, respectively. It is likely that the [O~{\sc i}] lines
are of sky origin, while the [N~{\sc ii}] lines derive from the nebula.
If so then the nebular expansion velocity, derived from HWHM of the [N~{\sc ii}] line is 26~km~s$^{-1}$.
This is higher than 4.6 km~s$^{-1}$, the expansion
velocity derived by Pollacco et al. (1990) from the nebular line split. Although the line splitting
in their high resolution spectra is labeled as {\it possible}, due to a low signal-to-noise ratio,
their data exclude an expansion velocity as large as 26~km~s$^{-1}$. 

Other nebular lines in the LHJ94 spectrum of V348 Sgr include [N~{\sc ii}] $\lambda$6583, 
blended with stellar C~{\sc ii} (M2).
[O~{\sc ii}] $\lambda$$\lambda$3726,29 lie outside its spectral range, 
while [S~{\sc ii}] $\lambda$$\lambda$6717,31 are blended with stellar
C~{\sc ii} and O~{\sc ii} lines. 
The issue of H$\alpha$ has been discussed in Sec.~\ref{sec:stell_par}. The nebular contribution is likely to contaminate 
substantially the stellar emission. There is no doubt, however, of a stellar contribution,
since a blue shifted P-Cygni profile is evident in the LHJ94 spectrum.

Considering
the similarity of the spectra of V348 Sgr and HV~2671, one might wonder whether HV~2671 also has an ionized PN. 
However, the LHJ94 spectrum of V348 Sgr shows the difficulty of detecting the PN
when the star is at maximum light. Detecting the PN at maximum light {\it and} low resolution might therefore
be impossible.

Inspection of our low resolution spectrum of V348 Sgr, as seen in Fig.~\ref{fig:spec1}, reveals that even the PN 
lines known
to be present in the LHJ94 spectrum, namely the [N~{\sc ii}]
pair at 6548 and 6583~\AA , are hard to identify (the former is too weak and the latter blended with C~{\sc ii} lines).
[O~{\sc i}] $\lambda\lambda$6300,6363, observed in the LHJ94 spectrum, although suspected to be
of sky origin, are not observed at all in the Calar Alto low resolution spectrum.                
[S~{\sc ii}] $\lambda\lambda$6717,6731 are equally undetected. The only possible detection of a nebular line in the
low resolution spectrum of V348 Sgr are
[O~{\sc ii}] $\lambda\lambda$3726,3729,
but there is no evidence that the same feature is present in the spectrum of HV~2671 (although the blue part of
the spectrum suffers from a very poor signal-to-noise ratio).
In conclusion, there is no evidence that HV~2671 has an ionized nebula although the low resolution of its spectrum
should be blamed as the primary cause. 

\subsection{DY Cen \& MV Sgr}

Nebular emission lines of [O~{\sc i}], [S~{\sc ii}] and [N~{\sc ii}] are detected in the
spectrum of DY Cen (Rao et al. 1993), but are much narrower than those detected in the spectrum of
MV Sgr. 
From DY Cen's [S~{\sc ii}] lines at 6717 and 6731~\AA\ 
Rao et al. (1993) derive $N_e$ = (450$\pm$100)~cm$^{-3}$ (closer to the value for V348 Sgr than that for
MV Sgr), for a temperature range $T_e$ = 5000--20\,000~K. They estimate the
electron temperature to be $<$10\,000~K from the absence of [N~{\sc ii}] $\lambda$5755. The fact that the
lines in their spectrum are not extended leads to the conclusion that the nebula around DY Cen has a diameter of
less than 2~arcsec. This is considered peculiar since the nebula has low electron density and, at a distance of 4.8~kpc, should
have an extent 
of 7~arcsec (for $T_{BB}$ = 20~kK and $L$ = 2300~L$_\odot$). However, we have noted (Sec~\ref{sec:stell_par}) 
that DY Cen is likely to be further 
away
than 4.8~kpc.

In the spectrum of MV Sgr, nebular lines of [S~{\sc ii}] and [O~{\sc i}] were first seen by Herbig (1964).
Pandey et al. (1996) observed nebular lines such
as [N~{\sc ii}] and [O~{\sc i}] with widths and shapes similar to
those of the allowed lines  (i.e. $\sim$100~km~s$^{-1}$). 
In addition, they used the forbidden line fluxes to derive the electron density, 
$N_e$ = (3--7)$\times$10$^{6}$~cm$^{-3}$, and electron temperature,
$T_e$ = 8500--20\,000~K. 
We confirm the detection of broad [N~{\sc ii}] and [O~{\sc i}] lines in the 1995
spectrum of MV Sgr
(the [O~{\sc i}] lines at 6300 and 6363~\AA\ also have a narrow sky component). Due to their widths, one has to deduce that
they are forming possibly in a tenuous outer stellar envelope rather than
in a proper PN. The high electron density in particular confirms that the nebula around MV Sgr is unlike
a typical PN (although very young PNe like Hb~12 or IC~4997 have densities of 10$^{5.6}$ and 10$^{5.3}$~cm$^{-3}$,
respectively, Pottasch 1984) and is more likely to be the outer envelope of the star.
Forbidden lines were also found in the cool RCB stars by Herbig
(1949), who saw [O~{\sc ii}] $\lambda$3727 in the spectrum of R~CrB. More recently, broad
emission lines of [O~{\sc i}], [N~{\sc ii}] and [S~{\sc ii}] were seen during the
1992 decline of V854 Cen.  In the 1995-96 decline of R~CrB, Rao et al.
(1999) found both narrow [C~{\sc i}], [O~{\sc i}], [Ca~{\sc ii}] and [Fe~{\sc ii}] lines and broad
[Ca~{\sc ii}], [N~{\sc ii}] and [O~{\sc ii}] forbidden emission lines.                                                       

The issue of the presence and characteristics of nebulosities around
RCB stars and other 
H-deficient stars like 
extreme helium stars or hydrogen-deficient carbon stars, 
is paramount to solving the question of their origin. Beside the evidence for ionized shells around
the three hot RCB stars, we know that
UW Cen has a neutral nebula that is about 15~arcsec across (Pollacco et al. 1991; Clayton et al. 1999). R~CrB has a 
nebula only seen with IRAS, several arcminutes across (Gillett et al. 1986). V854
Cen has a 5-arcsec nebula seen in C~{\sc ii} $\lambda$1335 (Clayton \& Ayres 2001). No other RCB star is
known to have a shell of any kind. This has always been a major stumbling block to the theory that RCB and 
central stars of PN are closely related. In theory, if RCB derive from a born-again type scenario it is possible 
that their PN have long recombined.

\section{The light-curve properties}
\label{sec:lightcurve}

RCB stars are defined by their light-curve properties, where deep
declines are followed by recovery in a random fashion. However the range of 
light-curve activity exhibited by RCB stars is extremely varied. 
Two characteristics that have emerged so far are that (a) RCB stars produce
dust in `puffs' near the stellar surface and then blow them away and (b) that the dust production
is related to the pulsation cycle of the stars (Clayton 1996 and references therein). 
On the other hand we still wonder: is the dust
produced preferentially on an equatorial region? How can dust form near the surface of a 6000K-20000K-star?
Theoretical work has gone a long way to explaining dust formation in hostile environments 
(Woitke, Goeres \& Sedlmayr 1996), but problems remain. In Table~\ref{tab:mags} 
we list the maximum light properties of the four hot RCB stars.
We present their light-curves from 1992 to 2000 in Fig.~\ref{fig:light}. In Sec.~\ref{ssec:maximum_light},
we discuss the long term behavior of the maximum brightness of RCB stars.

V348 Sgr was discovered to be variable by Woods (1926) and suggested to be an RCB star by Parenago (1931).
The light-curve behavior is summarized by Herbig (1958), Hoffleit (1958), 
Heck et al. (1985) and the AAVSO. 
There are some photometric data as far back as 1900 and more or less complete coverage 
from 1928 to the present. 
V348 Sgr is the second most active RCB star known (after V854 Cen).
It has an average time between declines of 560 days (Jurcsik 1996).
Between 1954 and 1981, V348 Sgr was never at maximum light for a full year
(Bateson 1982).                                                                                                                     

HV~2671 has little photometric coverage before the MACHO era (Alcock et al. 2001).  
It was discovered as a Harvard Variable in the
early 1900's with a variation of 1.8 mag. It was seen in a deep decline in 1967 by Kurochkin (1992). 
In the MACHO data, shown in Fig.~\ref{fig:light}, 
which cover about 7 years, HV~2671 has two deep declines. 

DY Cen had four declines recorded between 1897 and 1927 but 
none since, although
the photometric coverage has not been continuous (Hoffleit 1930; Bateson 1978; 
Pollacco \& Hill 1991; Jurcsik 1996; AAVSO). 
The Royal Astronomical
Society of New Zealand (RASNZ) and the 
AAVSO have been following DY Cen since 1960
without any declines being detected (Bateson 1978). 
DY Cen has a relatively high hydrogen abundance much like the cool RCB star
V854 Cen. But unlike that very active 
star, DY Cen seems
to go against the trend 
which relates high decline activity with high hydrogen abundance (Jurcsik 1996).

Similarly, MV Sgr was 
seen to decline at least twice (1930 and 1945) but hasn't 
had a large documented decline since then (Hoffleit 1959). 
Like DY Cen, the photometric coverage for MV Sgr has not been 
continuous and there have been some reports of small declines (Jurcsik 1996; 
Landolt 2001 priv. comm.). 
MV Sgr has been followed continuously by the AAVSO
since 1987 without having any discernible declines. 
However,
at near-IR and IRAS wavelengths, MV Sgr and DY Cen show similar 
excesses to the much more active V348 Sgr, indicating 
the presence of warm dust (Kilkenny \& Whittet 1984; Walker 1985; Rao \& Nandy 1986). 
So MV Sgr and DY Cen are likely to be producing dust but not along our line of sight to the star.

The range of decline activity seen in the hot RCB stars mimics that seen in the 
cooler RCB stars (Clayton 1996). 
The historical light-curve of R CrB itself, which now stretches 
back over two hundred years, shows that there are periods of up to 10 years 
where the star had no decline and other periods where it has had several 
declines in one year (Mattei, Waagen, \& Foster 1991).

\subsection{Long-term behavior of the maximum brightness}
\label{ssec:maximum_light}

RCB stars are true irregular variables, where the visual brightness of the star declines
every time that dust forms near the stellar surface, in the line of sight to the 
observer. The star returns to maximum light when the dust in the line
of sight disperses. Therefore, the maximum light is thought to be representative of the intrinsic stellar
brightness (once circumstellar and interstellar reddening is taken into account). 
In this Section, we discuss {\it secular} changes in the
the maximum-light brightness (i.e. the stellar brightness {\it after} it recovers from a decline),
and investigate to what they might be due: 
temperature or luminosity changes, or a change in the circumstellar reddening.

The maximum-light brightness evolution for three of the four hot RCB stars (HV~2671 has
little coverage before 1992), is summarized in
Table~\ref{tab:mags} and Fig.~\ref{fig:light2} 
(light-curves are {\it not} plotted in Fig.~\ref{fig:light2}, only
the maximum light estimates are plotted). Three types of magnitudes are
plotted in Fig.~\ref{fig:light2}, photographic magnitudes ($m_{pg}$), measured
by eye or with an iris photometer on blue plates, visual magnitudes ($m_{vis}$),
measured by eye through a telescope, and Johnson $V$ photoelectric magnitudes.
The transformation of $m_{pg}$ to Johnson $B$ magnitudes can be found in Arp
(1961) and Pierce \& Jacoby (1995). For the three stars, the maximum-light
colors are somewhat variable. For V348 Sgr, $B-V$ varies between 0.32 and 0.40~mag;
In the case of DY Cen, $B-V$ varies between 0.31 and 0.37~mag. MV Sgr varies between 0.22 and 0.28~mag
(Landolt 2001 priv. comm.; Pollacco \& Hill
1991). For all these values, the transformation from $m_{pg}$ to Johnson $V$ 
($V-m_{pg}$ = 0.17 - 1.09$\times$($B-V$)) is small.  For the $B-V$ colors of the hot RCB stars,
$V-m_{pg}$ will range from --0.1 to --0.2~mag.
The uncertainties in
individual $m_{pg}$ values can be seen in the scatter in
Fig.~\ref{fig:light2}. Also, Schaeffer (1994) collected a large number of
$m_{pg}$ magnitudes by different observers for SN 1937C including
measurements by
Hoffleit from similar plates.  These indicate that $m_{pg}$ measurements have an
uncertainty of $\pm$0.2-0.3~mag.  There is no discernible difference
between the
Johnson $V$ and $m_{vis}$ measurements in the data plotted in
Fig.~\ref{fig:light2}.
A comparison of $V$ and $m_{vis}$ resulted in the following transformation,
$m_{vis} - V$  = 0.21$\times$($B-V$) (Stanton 1999).
So for $B-V$ = 0.3~mag,  $m_{vis} - V$ = 0.06~mag. i.e. the correction is small.
The photographic and visual magnitudes have been transformed to V magnitudes in Fig.~\ref{fig:light2}.

In the first half of the 1900's, V348 Sgr would reach a maximum brightness of
$m_{pg}\lesssim$11 mag (Hoffleit 1958), but more recently its maximum brightness is V$\sim$12~mag 
(Heck et al. 1985). The maximum-light brightness of V348 Sgr therefore faded $\sim$1.0 mag from 1930
to 1957. It seems to have been fading more slowly
since 1957. Its maximum-light brightness also seems to have increased
0.4~mag from 1921 to 1930, which differ by less than the admitted uncertainty.
Similarly, DY Cen is reported to be gradually fading (Bateson 1978; Pollacco \&
Hill 1991; Rao et al. 1993).  
It can be seen from Fig.~\ref{fig:light2}
that the maximum-light brightness of DY Cen faded 0.7~mag from 1975 to
1993, remaining constant since then.  
MV Sgr has today a maximum-light
brightness of V$~\sim$ 13~mag while in Hoffleit's data it is $m_{pg}\sim$12 mag.
The maximum-light brightness of MV Sgr therefore faded $\sim$1.0
mag from 1941 to 1980, but the lack of photometric coverage does not allow the
beginning of the fade to be accurately determined, although the fading was well
underway by 1963.  
The maximum-light brightness has been fairly constant since
then.  
HV~2671 has by far the least photometric coverage
of the four stars.  Based on the published data, median m$_{pg}$ = 16.4~mag pre-1940,
and B = 15.5~mag in 1967, little can be said about any possible evolution of its maximum brightness.

The fact that three out of the four known hot RCB stars seem to be fading slowly with time may indicate 
that stars in the hot RCB class are evolving rapidly. The change might in fact indicate either
an intrinsic fading of the star (due to an overall fading of the bolometric luminosity
or to a change of temperature)
or an increase in the circumstellar reddening. Taking first the case where the change in V brightness
is due to a change in temperature, we note that for constant $M_{\rm bol}$, $\Delta M_V \sim$1,
corresponds to a decrease of the BC by the same amount. Such a change would imply an increase by
about 6000~K in the 10-20kK range. (We use the bolometric corrections for normal supergiant stars 
[Allen, Astrophysical 
Quantities, 4$^{\rm th}$ Edition]. Although this is generally not appropriate for RCB stars, the fact that we 
are using relative values, and that we are not extracting precise quantities, make this method suitable.
Additionally the BC of Leuenhagen \& Hamann (1994) of --1.71~mag determined for the 20kK V348 Sgr, is similar
to the values listed in Astrophysical Quantities for a 17.6kK B2 supergiant of --1.58~mag).

An increase in the effective temperature between 11.1 and 17.6~kK
would be accompanied by a change in the star's $B-V$ by only
$\sim$--0.14~mag. If the change in $V$ by $\sim$1~mag were due to reddening, $\Delta$A$_v$ = 1 translates to
a $\Delta$($B-V$) $\sim$ 0.32~mag. Therefore the color can help discriminate between a temperature and a reddening change.
The data are fragmentary but they show no color trend:
DY Cen was observed to have $m_V$ = 12.39~mag and $B-V$ = 0.31~mag in 1972
and $m_V$ = 12.78~mag and $B-V$ = 0.33~mag in 1987 (Rao et al. 1993). 
So for DY Cen, the fading is inconsistent with a reddening increase, while it is
more consistent with a temperature increase by $\sim$6000~K. 

The maximum-light brightness and color of MV Sgr was observed to be
$V$ = 12.70~mag and $B-V$ = 0.26~mag in 1963 (Herbig 1964) and $V$ = 13.29~mag, $B-V$ = 0.262~mag in 2000
(average of 8 observations, Landolt 2001 priv. comm.). Its maximum-light
brightness faded by $\Delta V$ = 0.6~mag but its $B-V$ color remained
unchanged. If the fading was due to added circumstellar reddening then we
would expect $B-V$ to become redder by $\sim$0.2~mag. So like DY Cen, the MV
Sgr fading is inconsistent with a reddening increase.
V348 Sgr is
much more active than the other two stars and could therefore be a better candidate for a reddening increase,
rather than a change in temperature.
However we would still like to argue that V348 Sgr, too, has suffered an increase in temperature, rather than
a change in reddening. The evidence is circumstantial for this star, but it is strongly
suggestive. In Fig.~\ref{fig:light2} we compare the long term light-curves for the three hot RCB stars with those
of the very active {\it cool} RCB stars R~CrB and RY Sgr, for which we have light-curves going
back over 150 years (unfortunately there are no other RCB stars with such ample
light-curve coverage). The two cool RCB stars show no maximum light fading trend.

Finally, we should point out that the $V$ brightness data is also consistent with a decrease in the bolometric luminosity.
A $\Delta M_{\rm Bol}$ = 1.0~mag corresponds to a decrease in luminosity by a factor of 2.5. This
cannot be excluded, although this type of luminosity changes would be hard to interpret
in the light of single star post-AGB evolution.

\section{Secular spectral variability}
\label{sec:SecularSpectralVariabilityOfTheHotRCBStars}

Here we check whether the variability in luminosity is accompanied by
any type of spectral variability over the 
period of time for which archival data were available. 
We stress that the spectral variability we are discussing in this Section 
is not the spectral variability observed during light declines, but
rather a change in the maximum-light spectrum between two different epochs.

\subsection{Historical Variability}
\label{ssec:historical_variability}

Of the four hot RCB stars, two, V348 Sgr and MV Sgr have spectral data going back to the 1940s
(Tables~\ref{tab:spectra1} and \ref{tab:spectra2}). We obtained and scanned some of the plates 
(ticked in Tables~\ref{tab:spectra1} and \ref{tab:spectra2}),
to search for any signs of variability.
We note that on the plate scans only significant spectral
changes can be detected. A spectrum's continuum level and shape are not indicative of the true stellar continuum, so that
only the lines can be considered. We can therefore detect the presence or absence of emission lines, or a 
complete change of the line makeup of the spectrum. Changes in temperature, abundance or in wind strength
inducing even a 50\% change in our lines' equivalent widths might be missed.

The 1949, 1950, 1954 and 1957 spectra of V348 Sgr (range $\sim$3500-5500~\AA ) were compared to 
the modern digital spectra, degraded to simulate the resolution and quality achieved by the scans of the historical spectra. 
We find no significant sign of change. However, the red spectrum ($\sim$5500-7000~\AA )
from 1953 
exhibits a deep absorption line at about 5850~\AA . This corresponds exactly with a strong 
emission in the degraded modern spectrum. The emission is a blend of Ne~{\sc i} M6, and the red-most component of C~{\sc ii}
M22. The rest of the features correspond rather well, so that a wavelength error is excluded. It is unclear why only this
feature should exhibit change. Overall we can say that no significant changes have occurred to V348 Sgr between 1949
and 1995.

The 1958 spectrum of MV Sgr is, once scanned, completely useless. However we can compare the 1963 spectrum
(in the range 4500-7000~\AA ) with the 1973 spectrum and the degraded 1995 spectrum. From this comparison it can be said that there is no clear sign of
spectral variability between 1963 and 1995. 
We stress that it is impossible to compare line
strengths and line ratios, but that all the lines seem to be there. 

Finally we should point out that the stellar temperature change suggested in Sec.~\ref{ssec:maximum_light}, 
might not produce noticeable changes in the scans of early spectra. 

\subsection{Recent Spectral Variability}
\label{ssec:recent_spectral_variability}

Despite the lack of obvious spectral variability over the last half 
century, we found that the high resolution, high signal-to-noise digital spectra
of V348 Sgr and DY Cen, show evidence of variability over the last decade.

\subsubsection{V348 Sgr and HV2671}

We compared two spectra of V348~Sgr, the LHJ94 spectrum taken in 1987
and the Calar Alto spectrum taken in 1995. The different resolutions are accounted 
for
by re-binning the higher resolution spectrum (R = 55,000) to the lower resolution one (R$\sim$5000). 
In Fig.~\ref{fig:v348_var} we compare the two rectified spectra in the range 5840--5950~\AA . 
The emission lines have increased in strength between 1987 and 1995 and there is an
indication that the P-Cygni profiles have shifted to bluer wavelengths (e.g., He~{\sc i} $\lambda$5876). At this
resolution it is unclear whether the line splits have changed between the two epochs.
If confirmed by a third epoch of observations, these changes could be interpreted as an
increase in wind density and velocity.

We compared the two maximum-light spectra of HV~2671 after degrading the slightly higher resolution of the
1995 spectrum to match the 1997 spectrum.
The 1995 spectrum was taken in non-photometric conditions and its overall flux level
had to be multiplied by a factor of 20 to match the continuum level of the 1997 spectrum. 
Within the limitation of the resolution
as well as the low signal-to-noise ratio of the 1995 spectrum, we can say that no significant variation 
occurred in the intervening years, although the details of some of the emission line profiles are different.
With the current very limited data-set it is not clear whether this is a real effect.

\subsubsection{DY Cen and MV Sgr}

DY Cen was observed in 1989 and 1992 by Giridhar et al. (1996) and in 2001 (April and September)
by us. Unfortunately the two data-sets do not have an overlapping spectral range. 
A comparison of the 1989 and 1992 spectra led Giridhar et al. (1996) to conclude that 
1) the mean radial velocity of the absorption lines had changed between 42 and 29~km~s$^{-1}$; 2) Some C~{\sc ii}
lines (e.g. $\lambda$$\lambda$5535.4,5537.6 [M10], $\lambda$$\lambda$6095.3,6098.5,6102.6 [M24]) had blue-shifted absorption in 1989,
and red-shifted absorption in 1992; 3) H$\alpha$ acquired a blue-shifted 
absorption between 1989 and 1992; and 4) He~{\sc i} $\lambda$5876 acquired a blue-shifted absorption 
and a split emission line
in 1992, while in 1989 it had only a split absorption. We have re-examined these spectra and confirm the
behavior observed by Giridhar et al. (1996). 
In Fig.~\ref{fig:v348_dycen_comp} we show 
four lines which changed. 
The 1992 spectrum of DY Cen shows a stronger H$\alpha$
with a weak P-Cygni absorption, not unlike the spectrum of V348 Sgr,
while the 1989 spectrum 
shows only a slightly asymmetric emission. The He~{\sc i} line at 5876~\AA\ shows the
emergence of a blue-shifted absorption, while the core of the line is rising into a split emission (much like
the profile observed in the spectrum of V348 Sgr, plotted alongside for comparison, thin solid line).
Last, the C~{\sc ii} M2 doublet, which is in pure absorption in the 1989 spectrum
(note that the emission peak is the nebular [N~{\sc ii}] line at 6583~\AA ), is also acquiring blue-shifted absorption in
1992, while the absorption troughs are partly filled.
In Fig.~\ref{fig:ha_dycen} we show H$\alpha$ and
the two C~{\sc ii} components of the M2 doublet, in their respective radial-velocity-corrected
velocity space. Here, we can see how the newly formed P-Cygni
profiles match for the three lines, while the still weak emission components appear to have a split appearance. 

In the 2001 spectra, the C~{\sc ii} line $\lambda$4267 is a weak but 
very broad (590~km~s$^{-1}$; Fig.~\ref{fig:2001obs}) emission. Although this spectral range cannot be compared 
directly to the red spectrum of Giridhar et al. (1996), the presence of a broad emission line such as 
the one observed is {\it per se} an indication of wind activity. The line variability observed (the disappearance of
the narrow emission component in the September spectrum [Fig.~\ref{fig:2001obs}]), might be a short term variability
superimposed on a more general behavior. Once again, a third epoch of observations, overlapping in spectral
range with early spectra, needs to be taken to determine whether the wind of DY Cen is truly strengthening.

MV~Sgr was observed on several occasions (Table~\ref{tab:spectra1} and \ref{tab:spectra2}).
No variations were seen.  The maximum-light spectra
all look very similar.

\vspace{0.4in}
In conclusion, two out of the four known hot RCB stars might show signs of increased wind density, although
more observations are needed to confirm this trend. Moreover, we must beware of the fact
that additional, short term variability might be present (like in the case of DY Cen). 
If these changes were attributed to increased wind mass-loss, we
could propose that the
{\it luminosity} has increased since it is by far the largest cause 
of changes in the wind efficiency. On the other hand, this would be very surprising in view of the
post-AGB nature of these stars. We might therefore have to invoke an increasing temperature, 
and opacity (i.e., a higher metal-to-hydrogen ratio, brought about by mass-loss). 
Without a third epoch of observations it is speculative to make any further conclusions.

\section{Summary of hot RCB star characteristics}
\label{sec:summary}

Before attempting an interpretation of the evolutionary status of the hot RCB stars
we summarize the salient points that characterize the hot
RCB stars. A summary is also presented in tabular form in Table~\ref{tab:summary}.

\begin{enumerate}
\item
V348 Sgr and HV~2671 have 
almost identical spectra at low resolution, exhibiting only emission lines. DY Cen has mostly absorption lines
in 1989 while in 1992 it has developed a few more emission lines. 
MV Sgr has strong emission lines that belong mostly to heavy elements,
with only weak carbon and nitrogen emission and absorption. 
Solely on the basis of the spectrum one would group V348 Sgr and HV~2671 together,
leaving DY Cen and MV Sgr in two separate categories.

\item
The stellar temperatures are in the range 15-20~kK. This, and the light
variability, are the only truly common
characteristics of the hot RCB stars. The abundances span a wide range. 
V348 Sgr, and presumably HV~2671, are very carbon-rich, grouping closely with the members of the [WC] central
stars of PN (e.g. Leuenhagen et al. 1996; 
De Marco \& Crowther 1998,1999; De Marco et al. 2002), 
DY Cen and MV Sgr are composed mainly of helium, more like normal RCB stars (Asplund et al. 2000).

\item
The emission lines of V348 Sgr have increased in intensity between 1987 and 1996, 
and the P-Cygni profiles have become deeper and bluer. 
DY Cen has suffered a similar fate between 1989 and 1992 where some of its absorption lines have turned into emission and
developed a weak P-Cygni absorption. On the other hand MV Sgr does not appear to have changed. For HV~2671 we do not
have enough data to draw any conclusions.
This behavior, seen in V348 Sgr and DY Cen, could be consistent with strengthening of the wind 
and hence an increase in mass-loss, but a third epoch with overlapping
spectral range is necessary in order to validate this conclusion.
The absorption lines of DY Cen have changed in position between 1989 and 1992. This
behavior is highly suggestive of binarity (as first suggested by Giridhar et al. 1996).

\item
V348 Sgr and DY Cen (in 1992) exhibit split emission lines of He~{\sc i} and C~{\sc ii} (though H$\alpha$, were it not
contaminated by nebular emission, would also show the same split). Some of these emission lines also exhibit P-Cygni
absorption. MV Sgr also exhibits split emission lines, although
mostly of heavy and s-process elements. Split lines are consistent with an equatorial density enhancement
like those encountered in B[e] stars.

\item
Three hot RCB stars exhibit forbidden emission lines. For the fourth, HV~2671, we have insufficient data. 
The forbidden lines in V348~Sgr and DY Cen point to low electron 
densities ($\sim$400~cm$^{-3}$),
and relatively low expansion velocities (10-20~km~s$^{-1}$), while the forbidden lines in the spectrum of MV Sgr
point to high velocities ($\sim$100~km~s$^{-1}$) and high densities ($\sim$10$^6$~cm$^{-3}$). Additionally, V348 Sgr
has a visible nebulosity about $\sim$30~arcsec (Pollacco et al. 1990) 
in diameter. No resolved nebulosity exists for the
other stars. Ionized nebulae around relatively cool stars might be consistent with the possibility that these stars
might have been hotter in the past.

\item
The maximum-light $V$ brightness of three hot RCB has declined in the last 70-100 years. 
We favor a temperature increase of $\sim$5000~K, to interpret this fading. For DY Cen
and MV Sgr, this is confirmed 
by the lack of a color change, which would be expected in case of a reddening increase.
No color data exist for V348 Sgr. 

\item
All four stars have been classified as RCB stars on the basis of their large, random
declines in brightness. At present, V348 Sgr and HV~2671 are very active, while MV Sgr
and DY Cen have had no declines in 50 years and are among the least active RCB stars known.
IR excess data indicates, however, that new dust is present around all four stars.
\end{enumerate}

\section{Comparison with evolutionary models}
\label{sec:CommparisonWithEvolutionaryModels}

In this Section, we compare the properties of the four hot
RCB stars, summarized in Sec.\ref{sec:summary}, with the latest evolutionary
models (Herwig et al. 1997; Herwig 2000,2001; Bloecker 2001). 
We will exclude HV~2671 from this discussion in view of the scant information
we have about this star. However it is likely that this object groups together with
V348 Sgr.

{\bf V348 Sgr.}
Clearly, V348 Sgr distinguishes itself from the other two objects because of its high 
carbon abundance. Because of its prominent emission line spectrum, it has been also
been classified as a [WC12] Wolf-Rayet central star of PN. However, it does not fit 
seamlessly into the late type Wolf-Rayet star class (see the properties listed by 
Koesterke [2001]). V348 Sgr has a large, extended PN in 
contrast to most other stars of this group which have rather young (compact and high density) PNe 
(e.g CPD-56$^{\rm o}$8032 and He2-113 [De Marco et al. 1997], M4-18 [De Marco \& Crowther 1998]), 
although it is not dissimilar from the large
tenuous PN of the [WC11] star K2-16. 
No doubt, the [WC] class remains enigmatic,
but the very fact that it is an heterogeneous one (possibly comprising several undiscovered born-again stars), 
means that it can accommodate
V348 Sgr among its members. 
Thus, from the properties of the PN, V348 Sgr might in fact be the only late-type
[WC] star which owes its H-deficiency to a born-again evolutionary history.
However, 
V348 Sgr displays a suspiciously small oxygen abundance. If this star had initially
formed 
from a sun-like composition, then its initial oxygen abundance
would have been 
between 0.5 and 1$\%$ by mass. Herwig (2000) presented AGB models
which include additional mixing processes resulting in abundances consistent with the
oxygen mass fractions observed in the majority of the [WC]-like and PG1159
stars. Basically, additional dredge-up induced by convective
overshoot leads to a higher oxygen abundance in the AGB intershell. In the
course of the mixing (and possibly burning processes) associated
with the born-again evolution, the intershell material, which contains
significant amounts of carbon, come to the surface. If this mechanism
is responsible for the hydrogen-deficiency and the huge carbon
abundance of V348 Sgr ($41\%$), then we would expect a significant amount of 
oxygen, much higher then the observed upper limit of 1\%. 
[WC]-type (early and
late) and PG1159 stars have oxygen abundances in the range 5-15\%, consisten
with evolutionary models. Very low 
oxygen abundances are however not unheard of. One
other [WC]-late type star (IRAS21282+5050) and one PG1159 star (NGC7094)
have similarly low oxygen abundances. 
Therefore, although the old nebula and
the
H-deficiency support a born-again evolutionary origin for V348 Sgr, the very low
oxygen abundance casts serious doubts on this interpretation. 

Unfortunately, the
alternative, a WD-merger scenario, does not help us 
either. While it would be possible to accommodate a very low oxygen abundance,
this scenario could not account for the large
carbon abundance. 

{\bf DY Cen.} This star has characteristics in common with both V348 Sgr and MV Sgr. The
nebular properties are similar to those of V348 Sgr (a low PN expansion velocity and
a low PN electron density, generally compatible with a born-again
evolution). In addition this star shares with V348 Sgr signs of
increasing wind activity. However, the stellar chemistry is - as is the
case in MV Sgr - dominated by helium like
the cool RCB stars. {\it Helium predominance is irreproducible by any current single star
AGB or post-AGB evolutionary calculations}.
Another difficulty is the high carbon abundance (15 times solar).
Even if it were possible to uncover the 
helium-rich layer left over from H-shell burning, 
evolutionary calculations predict nitrogen, not carbon, as the dominant
element after helium. 
In addition, although the understanding of
the internal mixing processes has improved, uncertainties remain.
DY Cen is clearly the star in the hot RCB class, which is most likely to
be a progeny of the majority group of cool RCB stars. 

{\bf MV Sgr.} Due to the predominance of helium in the stellar atmosphere of 
MV Sgr, we might conclude that this
star is too a typical descendant of the cool RCB class. It shares
with the other two stars the decreasing visual luminosity, which we
interpret as increasing stellar temperature, indicating a leftward motion in the
HR diagram, consistent with a post-AGB star increasing in temperature. However, the CNO elements are
dominated by nitrogen (after helium) as expected for the result of hydrogen burning. The
interpretation that we do, in fact, see the helium-layer, resulting from
hydrogen-shell burning, is consistent with the absence of any hydrogen. However, it is
unclear how hydrogen was eliminated to uncover the helium layer:
the absence of hydrogen cannot result from hydrogen-ingestion during a very
late thermal pulse (VLTP): during a VLTP, hydrogen
is ingested into the $^{12}$C-rich layers of the intershell region. There the $^{12}$C
abundance is such that proton-capture transformation of $^{12}$C into
$^{13}$C dominates, and only a little $^{14}$N, if any, can form. MV Sgr has
a high expansion velocity, high density, hydrogen-rich nebula, inconsistent with the born-again scenario
unless deeper imaging detect a low-density extended PN around it, and/or the central nebulosity 
proves to be hydrogen-deficient.
This star
requires an evolutionary scenario capable of removing the remaining
envelope completely, but not through dredge-up, which
would leave some trace of hydrogen in the photosphere, or through hydrogen-ingestion
mechanism associated with the VLTP. Current stellar evolutionary models
cannot explain this star.

\section{Concluding remarks}
\label{sec:TheEvolutionaryHistoryOfTheHotRCBStars}

The stellar abundances seems to point to at least three classes for the four hot RCB stars!
Yet, other characteristics, though never entirely homogeneous, seem to draw these stars together.
One question is therefore raised, whether a post-AGB star, with, say, RCB-type abundances
can change significantly its photospheric abundances to become, say, a [WC]-like type star.
Such change, happening not during the AGB, but in the post-AGB phase, is not in agreement with
evolutionary calculations. However, in view of the strangeness of these stellar classes, 
it cannot be excluded a priori.
The suggestion that the spectrum of DY Cen is developing stronger emission lines,
and becoming similar to that of the 
chemically very different V348 Sgr, make this possibility appealing. Further
spectroscopy, to confirm a long-term change which is not either periodic or stochastic shorter term variability, 
is therefore eagerly sought. 

If the hot RCB stars turn out to be the progeny of the cooler ones, we might also wonder about
the possibility that the hot RCB stars constitute a link between the cool RCB stars and  
the chemically similar extreme helium stars. This was extensively discussed by Pandey et al. (2001). They
conclude that it is not straightforward to establish a link between the two classes, on the basis
of different heavy element abundances. Again, if we could open the possibility of significant abundance
changes, there might be a new open evolutionary channel {\it out} of the RCB class.
O(He) stars, which are hot H-deficient post AGB stars (four are known, of which two are associated with a PN;
Werner 2001) have been proposed as descendants of the RCB stars. However they might be too
hydrogen-rich (LoTr4 has 33\% hydrogen by mass; Rauch et al. 1996,1998) to make definitive conclusions.

Last, we have shown how all the hot RCB stars for which we have
high resolution data, show split lines. We have interpreted these lines as 
an equatorial 
bulge, similar to what is interpreted to be around B[e] stars, and {\it not} a
Keplerian disk, which might have pointed to binarity.
The origin of this bulge is not understood, but it might be a key in the
explanation of the dust-making activity of these stars, since it might present a denser,
more shielded environment where dust could condense. Understanding the spectral
variability properties
of the split lines is paramount in characterizing the evolution of the
equatorial enhancement, as well as how it relates to the pulsation variability observed
in other RCB stars.

Regarding the possibility that RCB stars derive from the merger of a helium and a CO WD,
we note that DY Cen is likely to be a binary. Although further observations are
sought, this fact alone would contradict a merger scenario, which is more likely to result in a single star
(Jefferey \& Saio 2002).

\acknowledgments
We thank Simon Jeffery for providing the high resolution spectrum of V348 Sgr. Wolf-Rainer Hamann is thanked for providing the
unpublished Calar Alto spectrum of V348 Sgr, while Kameswara Rao, Sunetra Giridhar are thanked for the
unpublished MV Sgr spectrum and for both the DY Cen spectra used in this analysis. 
George Jacoby is thanked for helpful conversations.
Thanks to Tony Misch and George Herbig for help in obtaining and scanning the plate spectra. 
Thanks to Janet Mattei and the AAVSO for the photometric data.
OD gratefully acknowledges financial support from
the Asimov Fellowship program.
FH would like to thank
D.A. VandenBerg for support through his Operating Grant from the Natural
Science and Engineering Research Council of Canada.


\clearpage

\begin{figure*}
\epsscale{0.8}
\caption{Spectral atlas for three hot RCB stars DY Cen (lower spectrum), MV Sgr (middle spectrum) and V348 Sgr (upper spectrum).
Lines are identified by the ion name and multiplet number (in brackets) except for helium where only the
multiplet number is used. For details of the spectra see text.}
\label{fig:atlas1}
\end{figure*}

\begin{figure*}
\epsscale{0.8}
\caption{Spectral atlas for the three hot RCB stars DY Cen (lower spectrum), MV Sgr (middle spectrum) 
and V348 Sgr (upper spectrum).
Lines are identified by the ion name and multiplet number (in brackets) except for helium where only the
multiplet number is used. For details of the spectra see text.}
\label{fig:atlas2}
\end{figure*}

\begin{figure*}
\epsscale{0.8}
\caption{Spectra of V348 Sgr (top) and HV~2671 (bottom) taken at maximum light. The 
spectrum of HV~2671 
has been shifted down by 1.0 flux unit. The spectrum of HV~2671 has been corrected for its 
RV of 258 km s$^{-1}$ (Alcock et al. 1996).}
\label{fig:spec1}
\end{figure*}

\begin{figure*}
\epsscale{0.8}
\caption{A selection of C~{\sc ii} and He~{\sc i} lines in the spectrum of V348 Sgr,
showing split profiles. The lines have been
converted to velocity space using their rest wavelengths and corrected for the radial velocity of 130~km~s$^{-1}$.
The vertical line which marks the position of the lines' troughs is at 0~km~s$^{-1}$.}
\label{fig:pedestal}
\end{figure*}   

\begin{figure*}
\epsscale{0.8}
\caption{Fits to the C~{\sc ii} $\lambda$4267 line (top) and the He~{\sc i} $\lambda$5876 line (bottom) 
in the spectrum of V348 Sgr. 
The solid line corresponds to the data, the dashed one to the fit. The dotted line in panels (b) shows the part of
profile that was eliminated before fitting. Panels (c) show absorption cloud fits to the difference between
fit and data in panels (b).}
\label{fig:hei5876}
\end{figure*}   

\begin{figure*}
\epsscale{0.8}
\caption{Left: comparison of the H$\alpha$ (solid line), C~{\sc ii} $\lambda$4745 (dotted line)
and He~{\sc i} $\lambda$5876 (dashed line) profiles for V348 Sgr. Right: 
determination of the stellar contribution (H$\alpha_{tot}^{obs}$ - H$\alpha_{neb}^{pre}$; dashed line) 
to the H$\alpha$ profile (H$\alpha_{tot}^{obs}$; dotted line). See text.}
\label{fig:ha_comp}
\end{figure*}        

\begin{figure*}
\epsscale{0.8}
\caption{Lightcurves of V348 Sgr, DY Cen, MV Sgr and HV~2671 from 1992 to 2000 (AAVSO, Alcock et al. 2001).}
\label{fig:light}
\end{figure*}

\begin{figure*}
\epsscale{0.8}
\caption{Maximum-light brightness measurements (not light-curves)
for the cool RCB stars R~Cr~B and RY Sgr, and for the hot RCB stars DY Cen,
MV Sgr and V348 Sgr from 1900 to 2000. Squares are photographic 
magnitudes, 
triangles are visual magnitudes from the RASNZ, asterisks are visual magnitudes from the AAVSO and crosses
are photoelectric photometry. All magnitudes were converted to the $V$ system.}
\label{fig:light2}
\end{figure*}

\begin{figure*}
\epsscale{0.8}
\caption{The degraded 1987 high resolution spectrum of V348 Sgr (LHJ94; thick solid line) compared to the
1995 intermediate resolution Calar Alto spectrum (dotted line).
Lines have increased in intensity between the two
epochs, while at the same time P-Cygni profiles have become more prominent and blue-shifted.  }
\label{fig:v348_var}
\end{figure*}        

\begin{figure*}
\epsscale{0.8}
\caption{A comparison of H$\alpha$, the C~{\sc ii} M2 doublet and He~{\sc i} M11, between the 1989 and 1992 
(thin dotted and thick solid lines, respectively) spectra of DY Cen. For H$\alpha$ and the He~{\sc i} lines 
the LHJ94 spectrum of V348 Sgr (thin solid line) is plotted alongside for comparison. Each of the four lines
developed a blue-shifted absorption in 1992, with increased emission,
indicating the emergence of a stellar wind between the two epochs.  }
\label{fig:v348_dycen_comp}
\end{figure*}       

\begin{figure*}
\epsscale{0.8}
\caption{The H$\alpha$ (solid) C~{\sc ii} $\lambda$6578 (dashed) and $\lambda$6583 (dotted) lines in the
1992 spectrum of DY Cen plotted in their
respective rest frames.}
\label{fig:ha_dycen}
\end{figure*}

\begin{figure}
\epsscale{0.8}
\caption{A comparison of the C~{\sc ii} $\lambda$4267 line from spectra taken in 2001 April (dotted line) and 
September (solid line), showing the line variability. The broad pedestal of the line, which
does not change between the two observations, is fit with a Gaussian curve (thin solid line).}
\label{fig:2001obs}
\end{figure}      

\clearpage

\begin{table}
\caption{Summary of the New and Published Visible Spectra of V348 Sgr and HV~2671}
\begin{scriptsize}
\begin{tabular}{lllllll}
&Date & Telescope&Wavelength& Resolution$^a$& Source&\\
\tableline
\tableline
V348 Sgr$^b$&&&&&&\\
\tableline
x&1949 July-August &Lick 36" refractor &4000-5000& 130 \AA~mm$^{-1}$&Herbig 1958&\\
x&1950 April       &Lick 36" refractor &4000-5000& 130 \AA~mm$^{-1}$&Herbig 1958&\\
 &1951 July        &Lick Crossley 36"  &3500-5000& 430 \AA~mm$^{-1}$&Herbig 1958&\\
x&1953 May         &Lick Crossley 36"  &3500-5000& 430 \AA~mm$^{-1}$&Herbig 1958&\\
 &1953 May         &Lick 36" refractor &4000-5000& 130 \AA~mm$^{-1}$&Herbig 1958&\\
 &1953 August      &Lick 36" refractor &red      & 200 \AA~mm$^{-1}$&Herbig 1958&\\
x&1954 June        &Lick Crossley 36"  &3500-5000& 430 \AA~mm$^{-1}$&Herbig 1958&\\
 &1955 July        &Lick Crossley 36"  &3500-5000& 430 \AA~mm$^{-1}$&Herbig 1958&\\
 &1956 June        &Lick Crossley 36"  &3500-5000& 430 \AA~mm$^{-1}$&Herbig 1958&\\
x&1957 May-June    &Lick 36" refractor &4000-5000& 130 \AA~mm$^{-1}$&Herbig 1958&\\
 &1958 June        &Palomar 200''      &3800-4800& 18  \AA~mm$^{-1}$&Houziaux 1968&\\
 &                 &                   &5000-6800& 27  \AA~mm$^{-1}$&Houziaux 1968&\\
 &1981 July-November&Lick 120"         &3500-8500& 550              &Dahari \& Osterbrock 1984&\\
 &1982 April       &CTIO 4m            &5000-6700& 25 \AA~mm$^{-1}$ &Dahari \& Osterbrock 1984&\\
 &1982 May-June    &Lick 120"          &3500-7100& 550, 1100        &Dahari \& Osterbrock 1984&\\
x&1987 April       &ESO 3.6m           &3900-4800& 55\,000          &LHJ94&\\
 &                 &                   &         &                  &Jeffery 1995&\\
 &1987 July        &AAT 3.9m           &3000-7400& 1100, 3000       &Pollacco et al. 1990&\\
x&1995 July        &Calar Alto 3.5m    &3780-6200& 5000             &Hamann priv. comm.&\\
x&1998 May         &INT 2.5m           &3500-6800& 1000             &This paper&\\
\tableline
HV~2671$^b$&&&&&&\\
\tableline
x  &1995 November&SAAO 74"&3500-5200&1400&Alcock et al. 1996&\\
x  &1997 December&SAAO 74"&3500-7500&1100&This paper&\\
\tableline
\end{tabular}
\end{scriptsize}
\tablenotetext {}{a. The resolution of photographic plates is in \AA~mm$^{-1}$. For
digital detectors, we give the resolving power, $R$, quoted by the sources. When the source provides only
a resolution in \AA ngstroms, without a reference wavelength, we assume 5500~\AA. \\ b. An ``x'' indicates that the
spectrum was used in the current analysis.} 
\label{tab:spectra1}
\end{table}

\begin{table}
\caption{Summary of the New and Published Visible Spectra of DY Cen and MV Sgr}
\begin{scriptsize}
\begin{tabular}{lllllll}
&Date & Telescope&Wavelength& Resolution$^a$& Source&\\
\tableline
\tableline
DY Cen$^b$&&&&&&\\
\tableline
 &1987 April&ESO 3.6m&4000-4900&27\,000&Jeffery \& Heber 1993&\\
 &1988 March&AAT 3.9m&3400-5200&4000&Pollacco \& Hill 1991&\\
x&1989 July&CTIO 4m&5500-7000&18\,000&Rao et al. 1993&\\
 &1990 July&CTIO 4m&4300-4500&27\,000&Rao et al. 1993&\\
x&1992 May&CTIO 4m&5500-7000&35\,000&Giridhar et al. 1996&\\
x&2001 April    &AAT 3.9m&3930-4800&35\,000&This paper\\
x&2001 September&AAT 3.9m&3930-4800&35\,000&This paper\\
\tableline
MV Sgr$^b$&&&&&&\\
\tableline
x&1958 May      &Lick Crossley 36"& 3500-5000& 430 \AA~mm$^{-1}$&Herbig 1964&\\
x&1963 June     &Lick 120"        & red      & 374 \AA~mm$^{-1}$&Herbig 1964&\\
 &1963 August   &Lick 120"        & blue     & 100 \AA~mm$^{-1}$&Herbig 1964&\\
x&1973 August   &Lick 120"        & 5500-7000& 34 \AA~mm$^{-1}$ &Herbig 1975&\\
 &1974 June     &Lick 120"        & 7500-8600& 34 \AA~mm$^{-1}$ &Herbig 1975&\\
 &1979 June     &AAT 3.9m         & 3400-3900& 8000             &Jeffery et al. 1988&\\
 &1980 May      &AAT 3.9m         & 4130-4500& 8000             &Jeffery et al. 1988&\\
 &1982 July     &AAT 3.9m         & 3900-8580& 14\,000, 4000    &Jeffery et al. 1988&\\
 &1985 April    &ESO 3.6m         & 3900-4800& 11\,000          &Jeffery et al. 1988&\\
 &1992 May      &CTIO 4m          & 5500-7000& 27\,000          &Pandey, Rao \& Lambert 1996&\\
x&1994 August   &AAT 3.9m         & 3900-4100& 8000             &Venn et al. 1998&\\
x&1994 September&SAAO 74''        &red       & 2000             &Venn et al. 1998&\\
x&1995 September&McDonald 2.7m    & 3720-8880& 55\,000          &Rao priv. comm.&\\
\tableline
\end{tabular}
\end{scriptsize}
\tablenotetext {}{a. The resolution of photographic plates is in \AA~mm$^{-1}$. For
digital detectors, we give the resolving power, $R$, quoted by the sources. When the source provides only
a resolution in \AA ngstroms, without a reference wavelength, we assume 5500~\AA. \\ b. An ``x'' indicates that the
spectrum was used in the current analysis.} 
\label{tab:spectra2}
\end{table}

\begin{table*}
\caption{Observational Data}
\begin{tabular}{cccccccc}
Star & $V$ & $E(B-V)$& Distance&$M_V$ & Helio. $RV$  & Declines  \\
    &(mag)& (mag)   & (kpc)   &(mag) & (km~s$^{-1}$)&           \\
\tableline
\tableline
V348 Sgr &12.0$^a$  & 0.45$^c$   & --     & --         & 130$\pm$5    & very active   \\
HV~2671  &16.1$^b$  & 0.15       & 50$^e$ &--3.0       & 259$\pm$31$^b$& active      \\
DY Cen   &13.0$^a$  & 0.5$^d$    & --     &--          & 22$\pm$2$^f$ & inactive      \\
MV Sgr   &13.4$^a$  & 0.43       & --     &--          & --95$\pm$5   & inactive   \\
\tableline
\end{tabular}
\tablenotetext{}{ 
a. AAVSO.
b. Alcock et al. 2001; 
c. Pollacco et al. (1990); 
d. Jeffery \& Heber (1993); 
e. Feast (1999); 
f. Giridhar et al. (1996);
}
\label{tab:basic}
\end{table*}

\begin{table*}
\caption{Reddening Toward V348 Sgr}
\begin{tabular}{lll}
Method& $E(B-V)$& Reference\\
\tableline
\tableline
UBV&0.9&Houziaux 1968\\
Nearby stars&0.4&Neckel \& Klare 1980\\
UBV&0.59&Heck et al. 1982\\
Balmer line decrement&1.4-1.5&Dahari \& Osterbrock 1984\\
C~{\sc ii} COG&0.9&''\\
2175 \AA bump&0.44&Heber et al. 1984\\
UBV&0.66&''\\
Balmer line decrement&0.3$\pm$0.2&Houziaux et al. 1987\\
Balmer line decrement&0.45$\pm$0.1&Pollacco et al. 1990\\
Stellar continuum&0.5&Leuenhagen \& Hamann 1994\\
Stellar continuum&0.63$\pm$0.02&Jeffery 1995\\
\tableline
\end{tabular}
\label{tab:red}
\end{table*}

\begin{table*}
\caption{Stellar Model Parameters.}
\begin{scriptsize}
\begin{tabular}{ccccccccccc}
\tableline
\tableline
Parameter                       &HV~2671   & V348 Sgr$^a$ & DY Cen$^e$&  MV Sgr           & RCB$^j$    & [WC]$^k$ & Sun$^l$\\
\tableline
$D$ (kpc)                       &  50.0    & 5.4     &    --       &  --                  & --         &-- &--\\
$E(B-V)$ (mag)                  &  0.15    & 0.5     &    0.5      &  0.38$^g$            & --         &-- &--\\
$m_V$ (mag)                     &  16.1    & 12.37   &    12.7     &  12.0$^g$            & --         &-- &--\\
$M_V$ (mag)                     & --3.0    & --2.84  &    --       &  --                  & --         &-- &--\\
$L$ (L$_\odot$)                 & 6025     & 5130    &    --       &  --                  & --         &-- &--\\
$BC$ (mag)                      &--1.71    &--1.71   &    --       &  --                  & --         &-- &--\\
$T_{eff}$ (kK)                  &  20.0    & 20.0    &    19.5     &  16.0$^g$            & --         &-- &--\\
$R$ (R$_\odot$)                 &  3.7     & 6.0     &    --       &  --                  & --         &-- &--\\
$\dot{M}$ (M$_\odot$~yr$^{-1}$) & --       & --6.5   & --          &  --                  & --         &-- &--\\
$\log(g$/cm~s$^{-2}$)           & --       & --      & 2.15        &  2.5$\pm$0.3$^h$     & --         &-- &--\\
$X_H$(\%)                       & --       &  4$^b$  & 2.4        &  $\sim$0$^{h,i}$     & 0-5     &$<$10  & 74.4\\
$X_{He}$(\%)                    & --       & 40      & 94.7      &  99.85$^h$           & 98.6      &40-50 & 25.0\\     
$X_C$(\%)                       & --       & 55      & 2.9         &   0.05$^h$           &  0.7     &40-50& 0.20\\
$X_N$(\%)                       & --       & 0.5-1$^c$& --          &   0.09$^h$           &  0.4      &0-1& 0.09\\
$X_O$(\%)                       & --       & $\le$0.01      & --          &  --                  &  0.2      &5-10& 0.80\\
$X_{Si}$(\%)                    & --       & $<$0.1$^c$& --        &   0.01$^h$           &  0.02     &--  & 0.09\\
Code                            & --       & non-LTE sp$^d$& LTE pp$^f$ & LTE pp$^f$      & non-LTE pp$^f$ &non-LTE sp$^d$&--\\
\tableline
\end{tabular}
\end{scriptsize}
\tablenotetext {}{
a. Leuenhagen \& Hamann 1994, Leuenhagen et al. 1996, Leuenhagen \& Hamann 1998;
b.
upper limit; see Sec.~\ref{ssec:stellar_parameters_v348};
c. Leuenhagen \& Hamann 1998;
d. Non-LTE, spherical geometry;
e. Jeffery \& Heber 1993;
f. LTE plane-parallel geometry;
g. Drilling et al. 1984;
h. Jeffrey et al. 1988;
i. Herbig 1964, Pollacco \& Hill 1990;
j. From Table 3 of Asplund et al. (2000);
k. Averages from Leuenhagen \& Hamann (1998) and Leuenhagen, Hamann \& Jeffery (1996);
l. Grevesse, Noel \& Sauval 1996.
}
\label{tab:model}
\end{table*}

\begin{table*}
\caption{Evolution of Maximum Brightness}
\begin{tabular}{llll}
Star            &Year           &Max Brightness &Reference\\
\tableline
\tableline
V348 Sgr        &1900-54        &$m_{pg}$=11.0          &1\\
                &1954-81        &$m_{vis}$=11.6-11.8    &2      \\
                &1973           &$V$=12.1               &3\\
                &1981           &$V$=12.3               &3\\
                &1990           &$V$=12.0               &4\\
                &1995           &$V$=11.9               &4\\
                &1998           &$V$=11.9               &4\\
                &1960-2001      &$m_{vis}$=12.0         &5\\
\tableline
DY Cen          &pre-1930       &$m_{pg}$=12.0          &6\\
                &1960-1977      &$m_{vis}$=12.4-12.7    &7\\
                &1972           &$V$=12.4               &8\\
                &1982-3         &$V$=12.5               &9\\
                &1987           &$V$=12.7               &10\\
                &1987           &$V$=12.8               &11\\
                &1990           &$V$=12.8               &4\\
                &1982-2001      &$m_{vis}$=12.7-13.0    &5\\
\tableline
MV Sgr          &pre-1928       &$m_{pg}$=12.1          &12\\
                &1924-49        &$m_{pg}$ = 12.0        &13\\
                &1963           &$V$=12.70              &14\\
                &1985           &$V$=13.34              &15\\
                &1980-2000      &$V$=13.0-13.4          &4\\
                &1987-2001      &$m_{vis}$ = 13.4       &5\\
\tableline
\end{tabular}
\tablenotetext {}{
1. Hoffleit 1958, 
2. Bateson 1982, 
3. Heck et al. 1985, 
4. Landolt 2001 priv. comm., 
5. AAVSO, 
6. Hoffleit 1930, 
7. Bateson 1978, 
8. Sherwood 1975, 
9. Kilkenny et al. 1985, 
10. Jeffery \& Heber 1993,
11. Pollacco \& Hill 1991,
12. Woods 1928, 
13. Hoffleit 1959,
14. Herbig 1964
15. Goldsmith et al. 1990.} 
\label{tab:mags}
\end{table*}

\begin{table*}
\caption{Fit Parameters for C~{\sc ii} $\lambda$4267 and He~{\sc i} $\lambda$5876} 
\begin{tabular}{cccccccc}
&&\multicolumn{3}{c}{ C~{\sc ii} $\lambda$4267} & \multicolumn{3}{c}{ He~{\sc i} $\lambda$5876 } \\
Test & Line Profile   & Center  & FWHM  &   EW & Center  & FWHM   &   EW \\
     &                & (km~s$^{-1}$) & (km~s$^{-1}$) & (\AA )& (km~s$^{-1}$) & (km~s$^{-1}$) & (\AA ) \\
\tableline
\tableline
 &   1     &      --15 &384  &  1.7 & --101 &  60 & --0.24\\
1&   2     &      {\bf--43} & {\bf47}  &  {\bf0.41}& {\bf--22}  &  {\bf46} &   {\bf0.61}\\
 &   3     &      {\bf 37} & {\bf74}  &  {\bf1.0} & {\bf  43}  &  {\bf49} &   {\bf1.2}\\
\tableline
2&   1     &      --26 &551  &  1.8 &  --96  &  49   &  --0.2\\ 
 &   2     &       {\bf 11} &{\bf 124}  & {\bf  2.2} &  {\bf 19} & {\bf  85}   &  {\bf  2.6}\\
\tableline
\end{tabular}
\label{tab:fit}
\end{table*}

\begin{table*}
\caption{Cloud Fits to the Absorption Lines Presented in Fig~\ref{fig:hei5876}, Panels (c).}
\begin{tabular}{cccc}
Profile                    &  $v$        & $\log(N$ & $b$ \\
                           &(km~s$^{-1}$) & /cm$^{-3}$)& (km~s$^{-1}$)\\
\tableline
\tableline
He~{\sc i} $\lambda$5875.6 & 8          & 12.3   & 18  \\
C~{\sc ii} $\lambda$4267.3 & --3          & 12.2   & 20 \\
\tableline
\end{tabular}
\label{tab:cloud}
\end{table*}

\begin{table*}
\caption{Summary of hot RCB star characteristics.}
\begin{tabular}{llll}
                       & V348 Sgr & DY Cen & MV Sgr \\
\tableline
\tableline
$n_e$(PN)        & low            & low           &high \\
$v_{exp}$(PN)    & low            & low           &high \\
split stellar lines   & yes            & yes           &yes\\
stellar          & [WC]-like      & He+C, no      &  H-burning ashes by\\
abundances       &                & sign of H-    &  CN-cycle, emission \\
                 &                & burning       &  of heavy elements\\
spectral         & strengthening  emission & strengthening emission & none\\
variability      & lines =increasing wind? & lines =increasing wind?&     \\
long-term maximum& decreasing                 & decreasing                &  decreasing \\
light variability& =increasing T$_{\rm eff}$  &  =increasing T$_{\rm eff}$& =increasing T$_{\rm eff}$           \\
decline activity & very high      & low           & low  \\
\tableline
\end{tabular}
\label{tab:summary}
\end{table*}

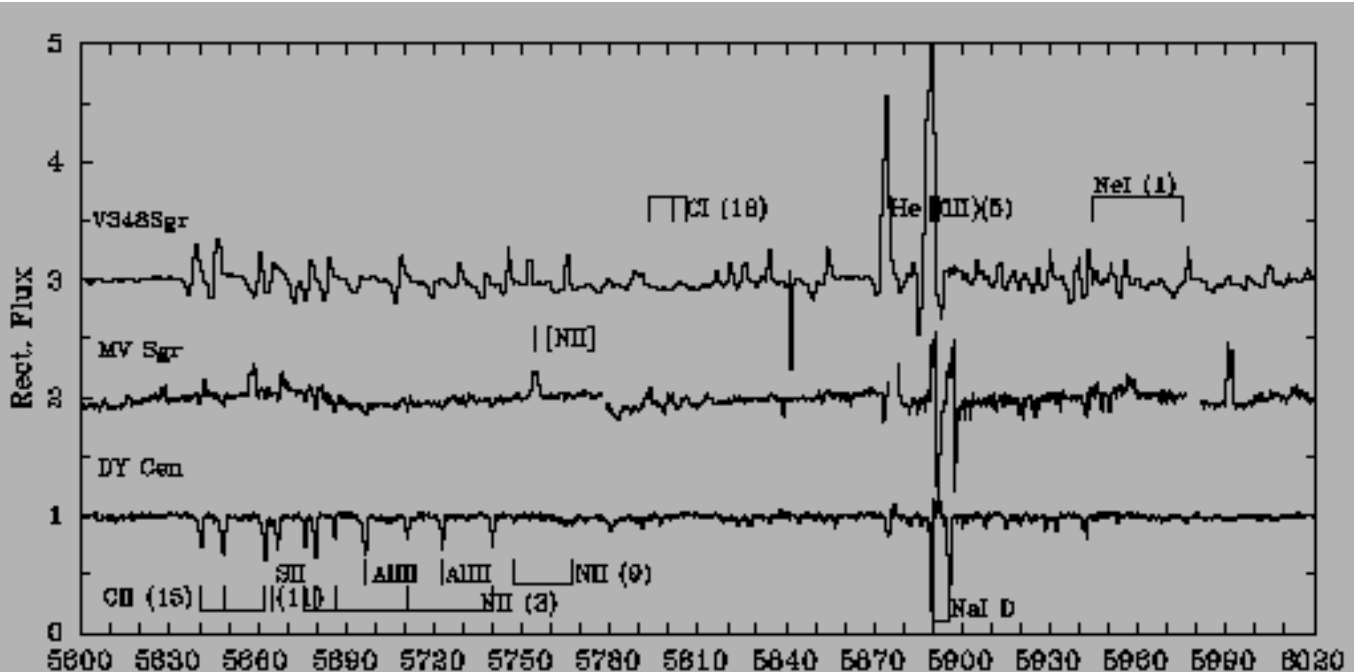
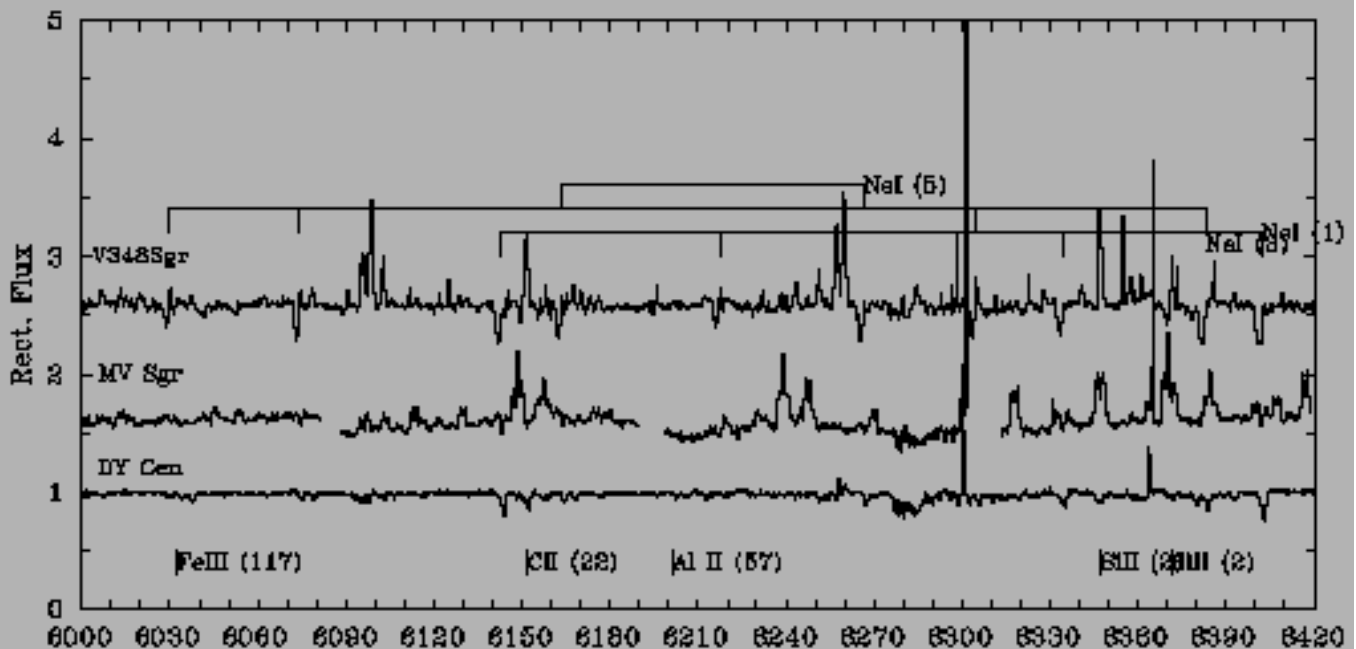
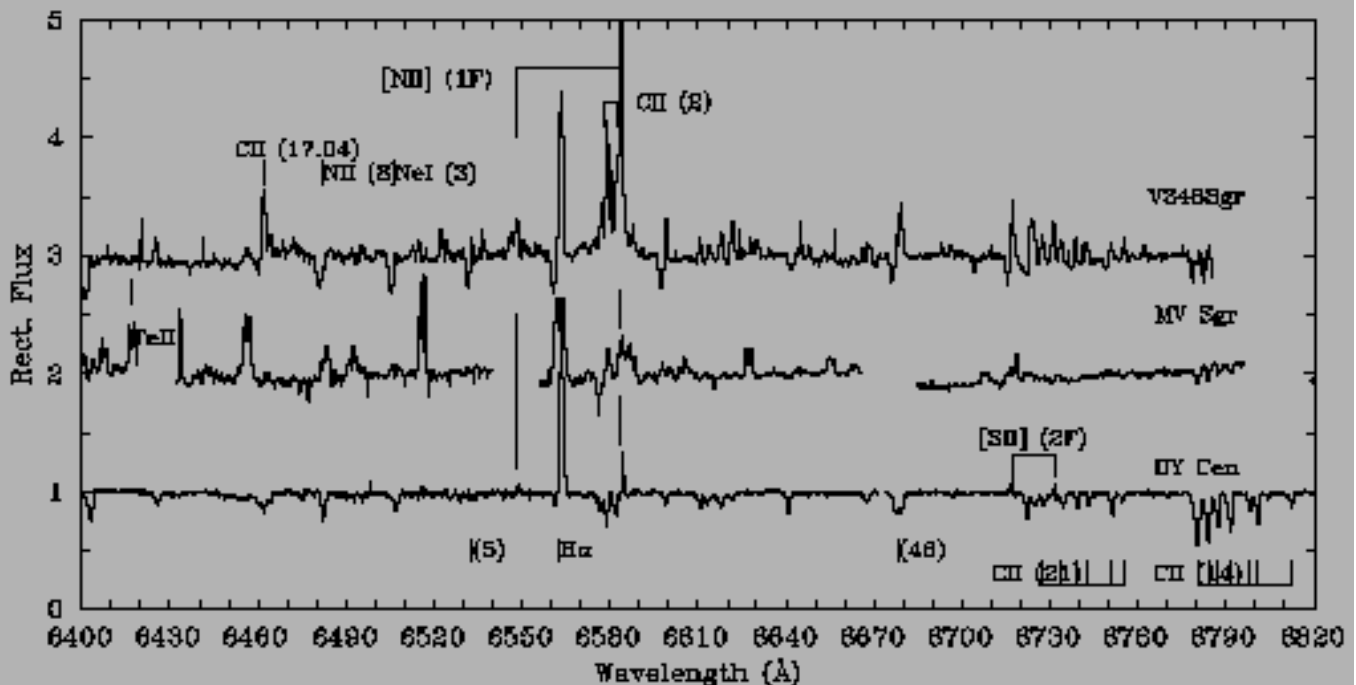
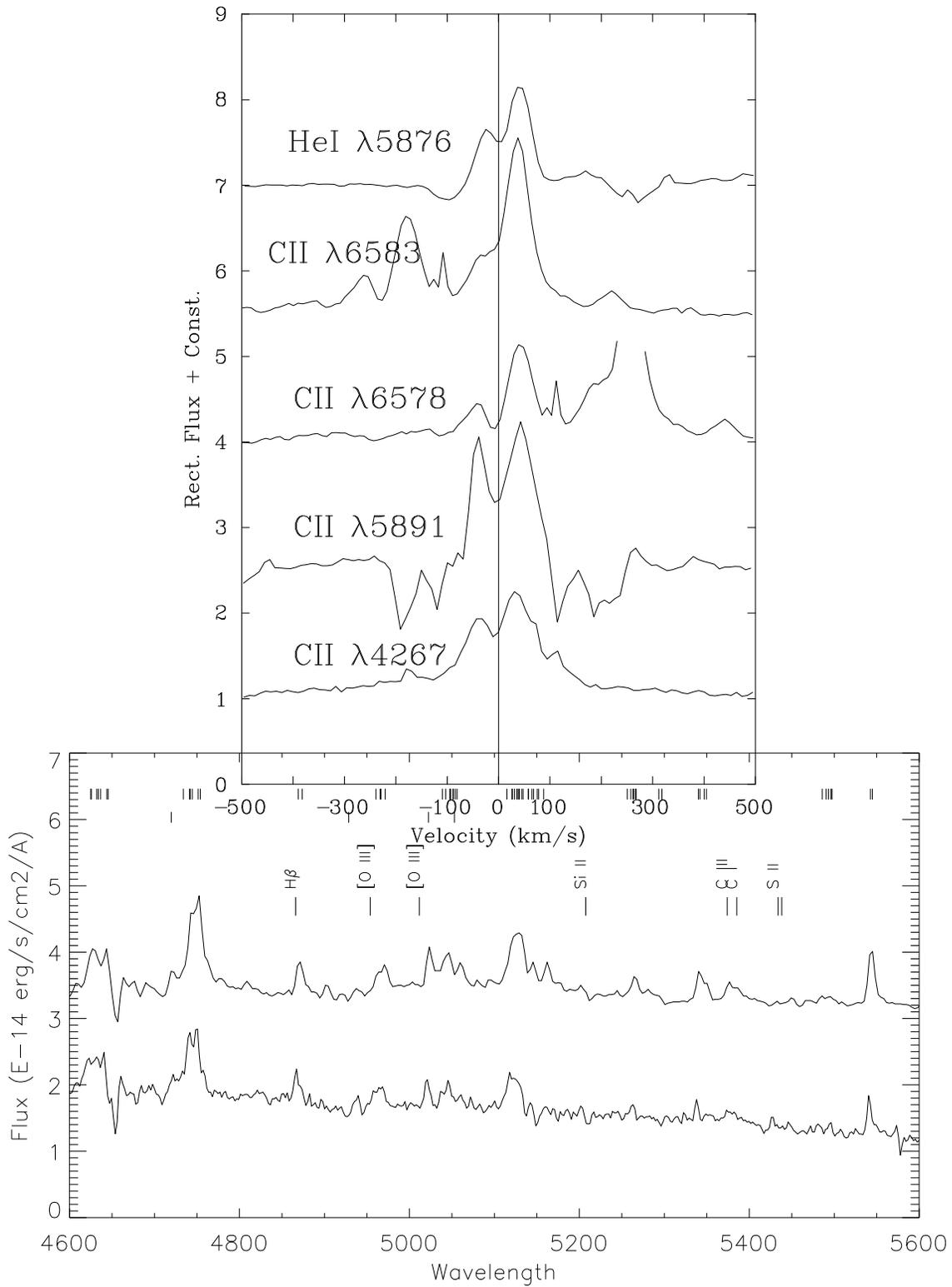
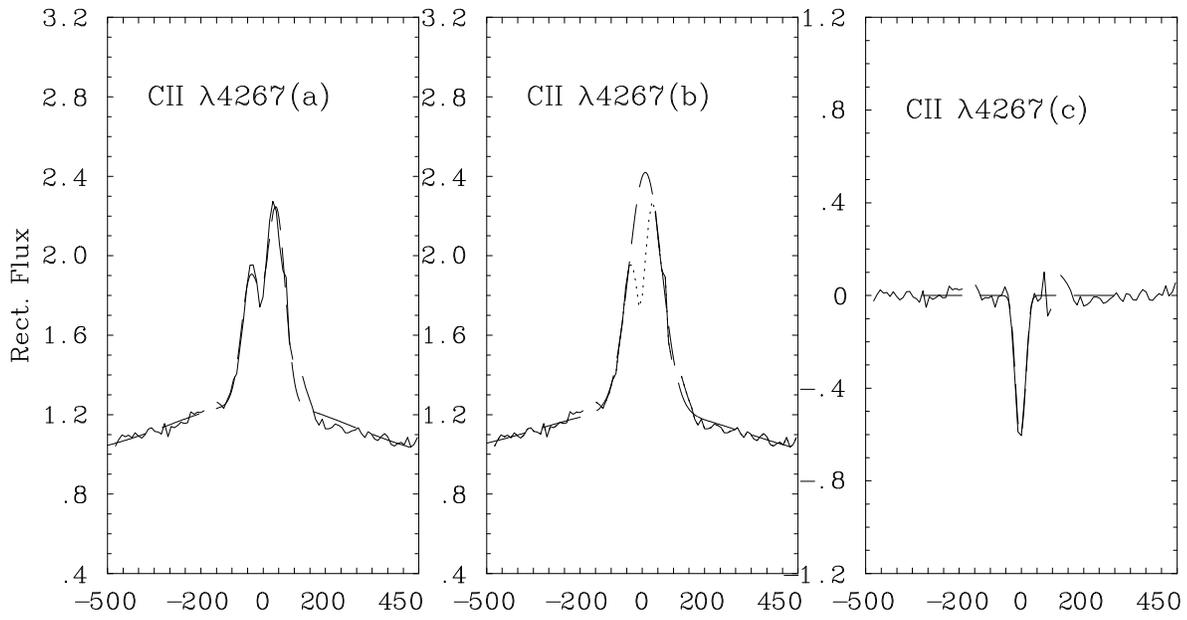
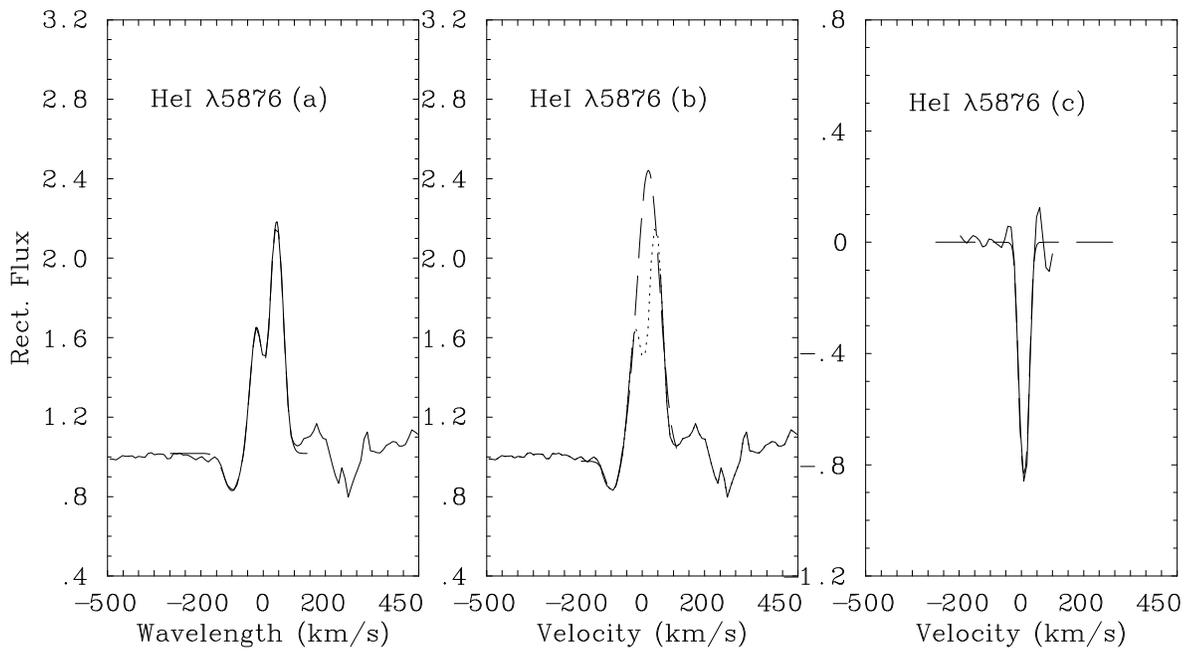
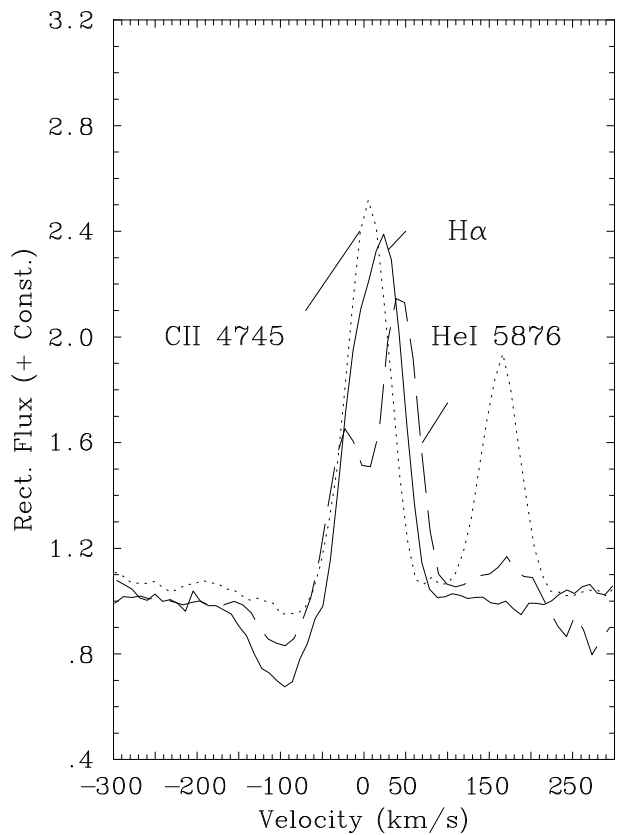
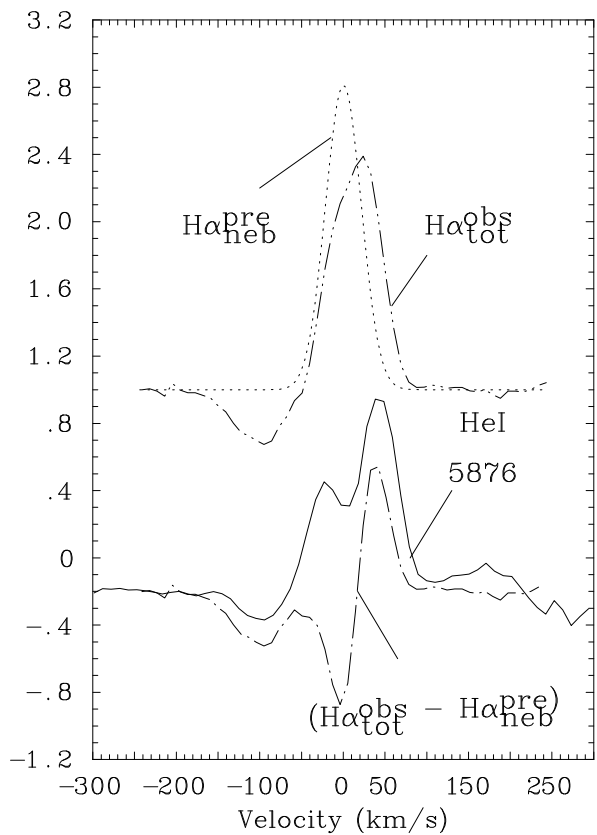
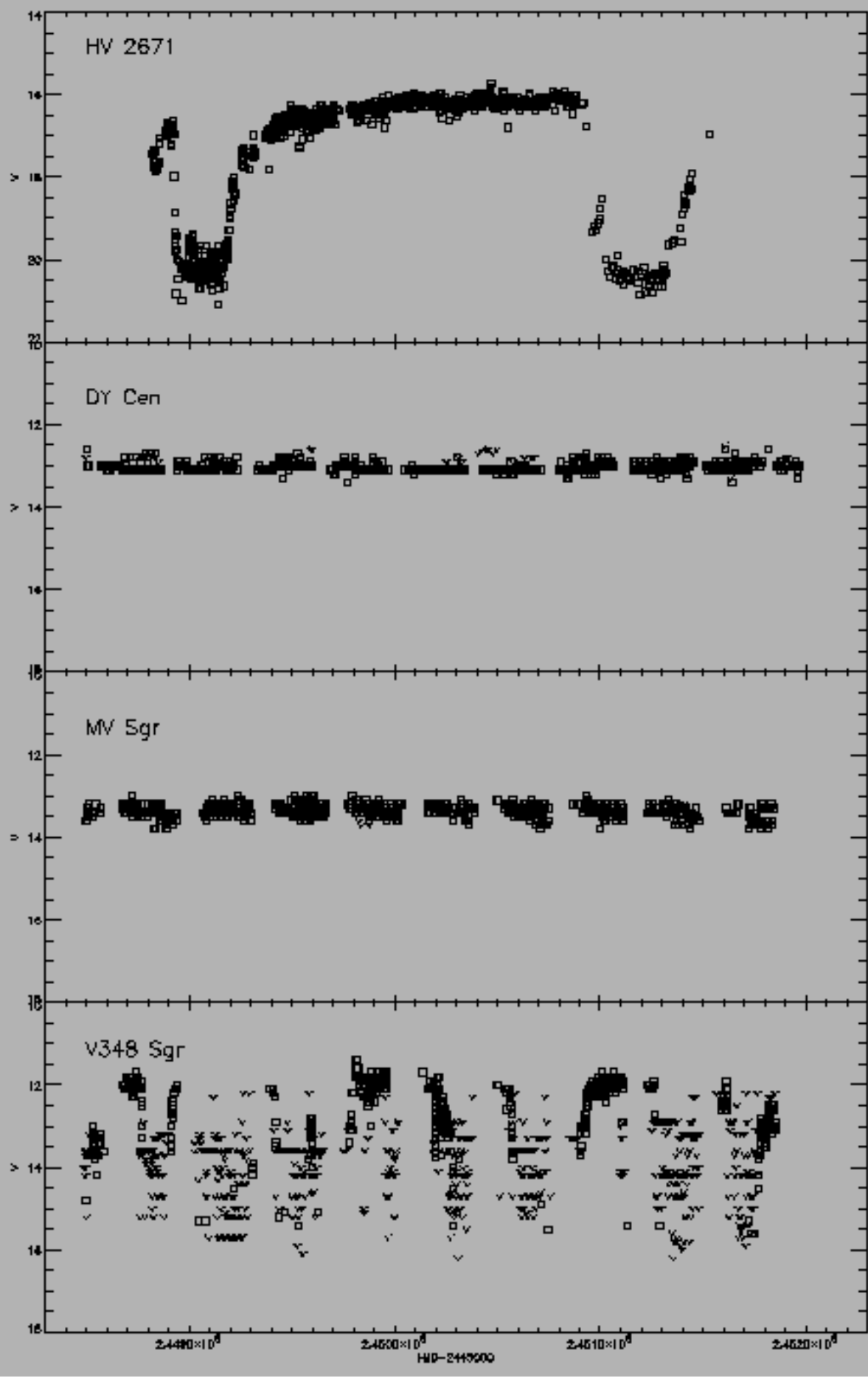
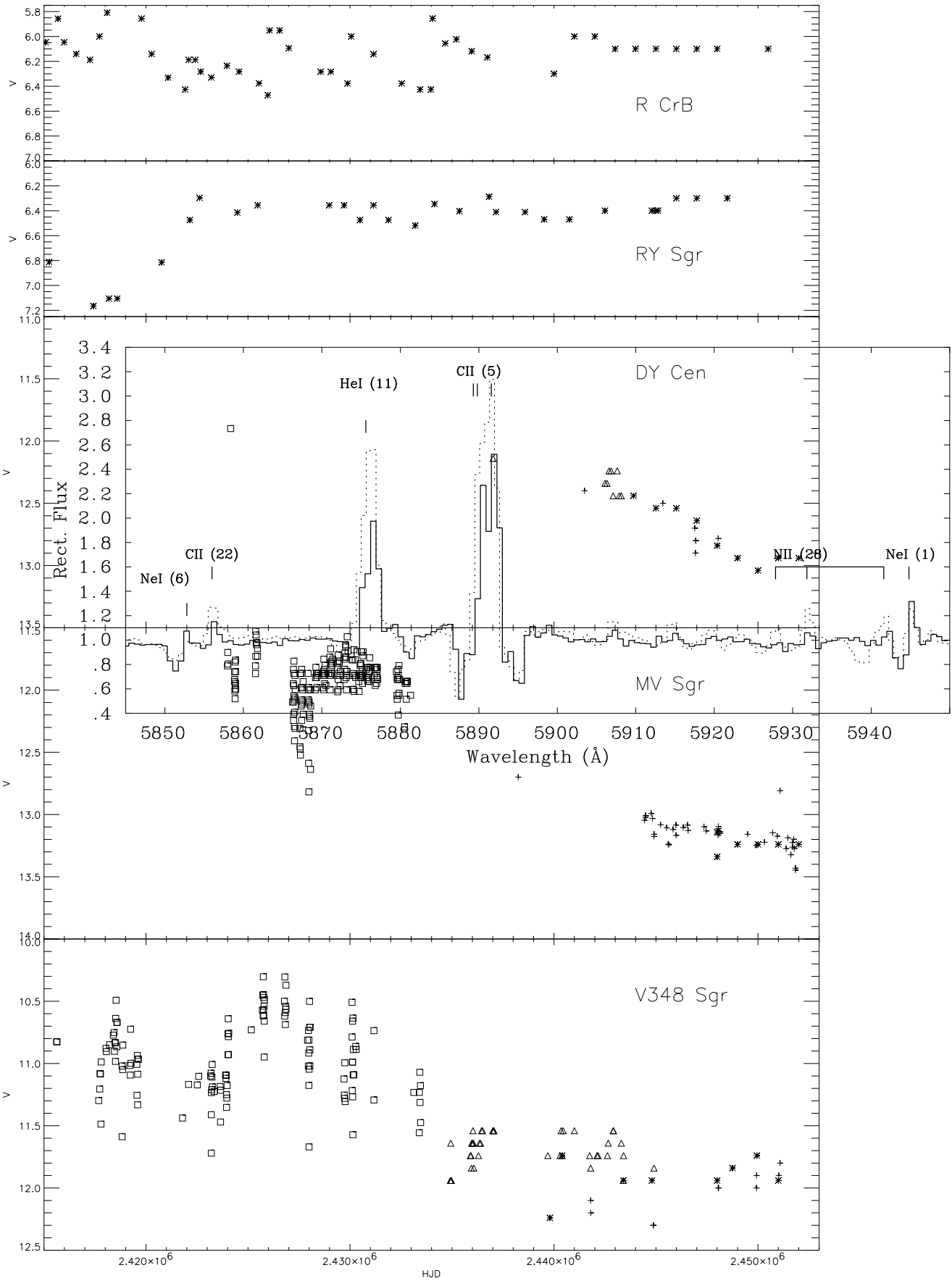
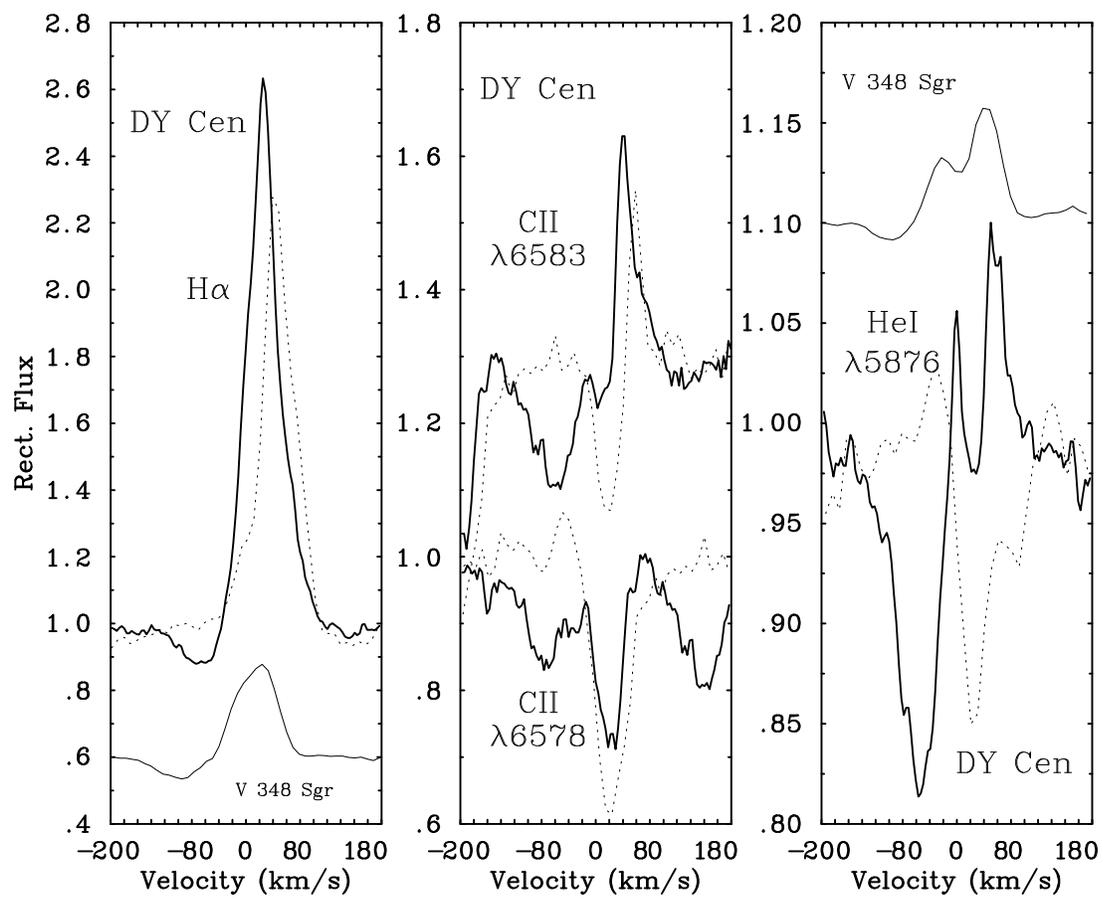
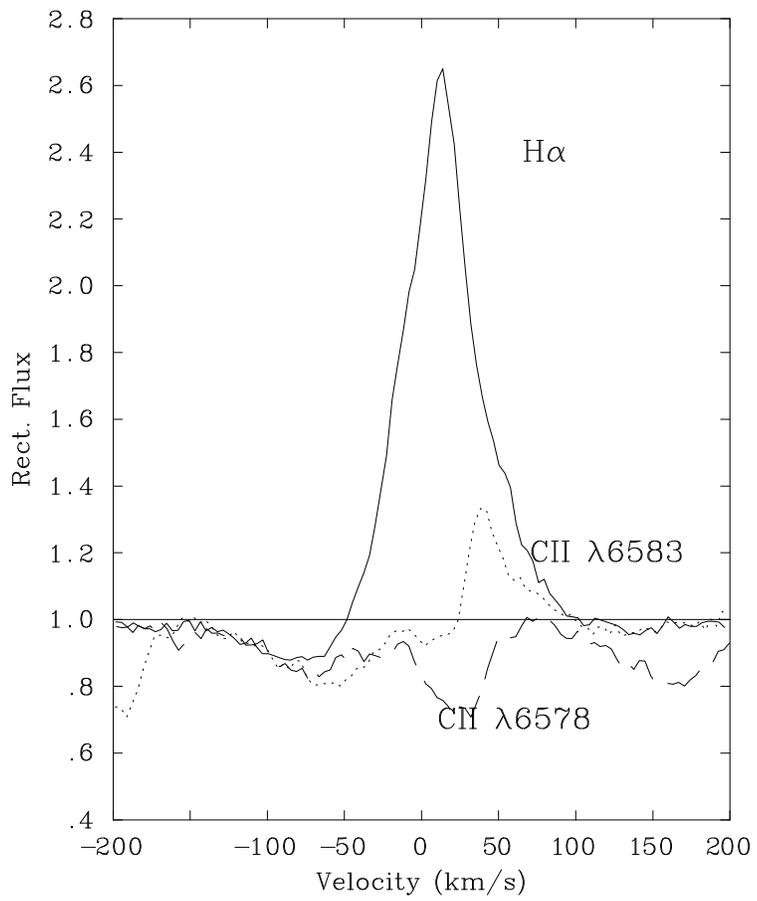
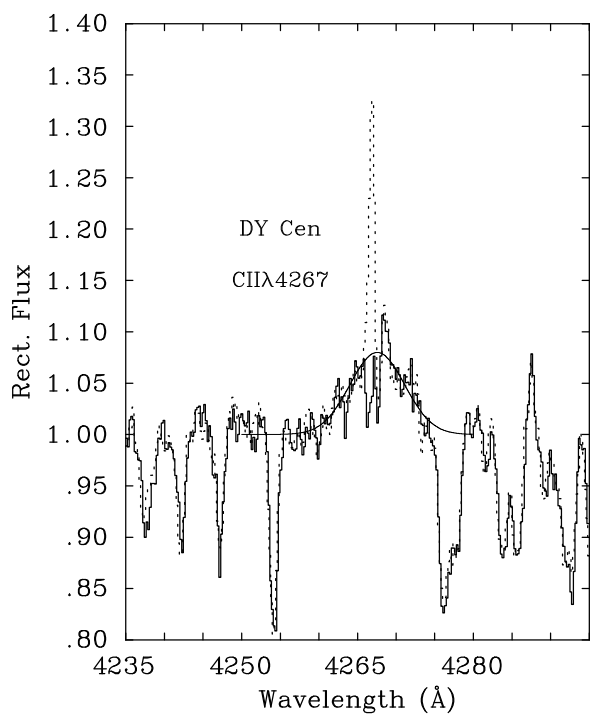
\end{document}